\documentclass[reprint, floatfix, superscriptaddress]{revtex4-1}
\usepackage{amsmath,amssymb}
\usepackage{graphicx,hyperref,framed}
\newcommand{\nn}{\nonumber}
\newcommand{\dd}{{\rm d}}
\begin{document}
\title{Dynamics in a one-dimensional ferrogel model: relaxation, pairing, shock-wave propagation}
\author{Segun Goh}
\email{segun.goh@hhu.de}
\affiliation{Institut f\"ur Theoretische Physik II: Weiche Materie, Heinrich-Heine-Universit\"at D\"usseldorf, D-40225 D\"usseldorf, Germany}
\author{Andreas M. Menzel}
\email{menzel@thphy.uni-duesseldorf.de}
\affiliation{Institut f\"ur Theoretische Physik II: Weiche Materie, Heinrich-Heine-Universit\"at D\"usseldorf, D-40225 D\"usseldorf, Germany}
\author{Hartmut L\"owen}
\email{hlowen@thphy.uni-duesseldorf.de}
\affiliation{Institut f\"ur Theoretische Physik II: Weiche Materie, Heinrich-Heine-Universit\"at D\"usseldorf, D-40225 D\"usseldorf, Germany}
\date{\today}
\begin{abstract}
Ferrogels are smart soft materials, consisting of a polymeric network and embedded magnetic 
particles. Novel phenomena, such as the variation of the overall mechanical properties by
external magnetic fields, emerge consequently. However, the dynamic behavior of ferrogels 
remains largely unveiled. In this paper, we consider a one-dimensional chain consisting of 
magnetic dipoles and elastic springs between them as a simple model for ferrogels. The model
is evaluated by corresponding simulations. To probe the dynamics theoretically, 
we investigate a continuum limit of the energy governing the system and the corresponding 
equation of motion. We provide general classification scenarios for the dynamics, 
elucidating the touching/detachment dynamics of the magnetic particles along the chain. 
In particular, it is verified in certain cases that the long-time relaxation corresponds 
to solutions of shock-wave propagation, while formations of particle pairs underlie the 
initial stage of the dynamics. We expect that these results will provide insight into the 
understanding of the dynamics of more realistic models with randomness in parameters and 
time-dependent magnetic fields.
\end{abstract}
\maketitle
%
\section{Introduction}
\label{sec:intro}
%
Ferrogels, magnetic elastomers, or magnetic gels are smart composite materials~\cite{Filipcsei2007}
the elastic properties of which are tunable by magnetic fields from 
outside~\cite{Ilg2013,Menzel2015,Odenbach2016,Lopez2016}. 
Novel characteristics originate from the composite nature of ferrogels 
which gives rise to a magneto-mechanical coupling between 
the embedded magnetic particles and the gel network~\cite{Frickel2011,Ilg2013,Allahyarov2014}. 
Such a magneto-mechanical coupling can be achieved by constraining the motion of magnetic 
particles inside pockets of the matrix~\cite{Frickel2011,Gundermann2014,Landers2015} 
or by directly anchoring the polymers to the surfaces of 
magnetic particles~\cite{Ilg2013,Frickel2011,Messing2011,Roeder2015}. 
Utilizing this characteristic, a variety of applications such as 
sensors~\cite{Szabo1998,Volkova2017}, actuators~\cite{Schmauch2017}, 
tunable devices~\cite{Deng2006,Sun2008}, 
medical scaffolds for tissue engineering~\cite{Bock2010,Zhao2011}, and biocomposites for 
controlled release~\cite{Muller2017} have been suggested.

Much effort has also been devoted to the theoretical understanding of the ferrogels.
Several routes are suggested and investigated to model these non-trivial materials. 
At the microscopic scale, bead-spring models to resolve the individual polymer chains 
connecting the embedded magnetic particles have been studied by means of computer 
simulations~\cite{Weeber2012,Ryzhkov2015,Weeber2015jcp,Weeber2015jmmm}. 
On the macroscale, hydrodynamic theories for ferrogels have been
developed~\cite{Jarkova2003,Bohlius2004}. Moreover, mesoscopic dipole-spring 
models~\cite{Annunziata2013,Cerda2013,Pessot2014,Pessot2015,Sanchez2015,Ivaneyko2015}
represent the polymeric matrix by spring-like interactions, while the magnetic 
particles are resolved and interact with each other via magnetic dipole-dipole interactions.
Alternatively, the elastic contributions can be described by matrix-mediated 
interactions~\cite{Biller2014,Biller2015,Puljiz2016,Puljiz2017,Menzel2017} 
in terms of continuum elasticity theory.

Recently, more attention begins to be paid to dynamic properties.
Analogously to the dynamics of magnetic colloidal systems~\cite{Heinrich2011,Yan2012,Alvarez2013,Dobnikar2013,Klapp2016}, 
new configurations or generally novel phenomena observable only in the dynamics are expected to emerge for ferrogels.
As an important example, the dynamic moduli/responses of ferrogels have been studied 
extensively~\cite{An2010,An2012,Tarama2014,Pessot2016,Nadzharyan2016,Pessot2018,Sorokin2017}.
To fully describe the dynamics far from equilibrium and the consequent transitions between 
qualitatively different configurations, it is necessary to address the approach and separation
dynamics of magnetic particles under changing mutual magnetic attraction and repulsion.
Indeed, the changes in particle distances are well known 
to affect the material properties of ferrogels. One of the most widely studied phenomena in 
this regard is the formation of chain-like aggregates which can cause drastic changes in the 
elastic properties of systems~\cite{Wood2011,Melenev2011,Zubarev2013,Romeis2016,Zubarev2016chain,
Lopez-Lopez2017,Gundermann2017,Schumann2017}.
It has been predicted theoretically that the detachment of magnetic particles in 
chain-like aggregates can give rise to the pronouncedly nonlinear, 
so-called superelastic stress-strain behavior~\cite{Cremer2015,Cremer2016}.
The formation of chain-like aggregates has been studied for various dipolar systems, for instance in combination with the Van der Waals interaction~\cite{vanRoij1996,Kwaadgras2013}.

In a theoretical perspective, the formation of compact chains under magnetic attraction can be 
viewed as a hardening transition~\cite{Annunziata2013}, if the particles can come into close 
contact. Steep changes in elastic properties can be attributed to the hardening due to 
virtual touching. It is worthwhile to note that the hardening 
transition implies a double-well structure in the energy. In other words, there exist two 
different equilibrium configurations, one of which corresponds to the contracted and the other
to the elongated systems. Such a configurational bistability, involving the rearrangement of 
the magnetic particles and the deformation of the gel network, has been widely discussed 
with different settings~\cite{Melenev2011,Annunziata2013,Biller2014,Biller2015,Zubarev2016hysteresis} 
and therefore seems to be a relatively universal feature. Moreover, one may expect 
that there exists a regime in between the equilibrium points where configurations become unstable. 
In short, a type of dynamic mechanism formally similar to spinodal decomposition
may play a significant role if attention is extended to dynamics~\cite{Chaikin2000}.

Spinodal decomposition occurs in various systems, representatively to the phase 
separation of binary systems described with the aid of the Cahn-Hilliard 
equation~\cite{Cahn1958, Bates1993, Elliott1987, Pego1989}. 
Recently spinodal lines were identified for systems of active 
Brownian particles~\cite{Bialke2013,Speck2014,Fily2014,Cates2010,Stenhammar2013,Wittkowski2014}.
The wetting phenomenon~\cite{Popescu2004, Dietrich2005} also provides an example with a 
special boundary condition due to the presence of a reservoir. However, there are technical 
difficulties related to the regularization of the problem~\cite{Hollig1983,Barenblatt1993} 
which corresponds to the unique characteristics of each system under consideration. 
In the case of the Cahn-Hilliard equation, for instance, the regularization term contains
the free energy cost due to the interface and therefore governs the coarsening dynamics in the 
long-time scale. Also see, e.g., Refs.~\onlinecite{Cates2010} and~\onlinecite{Wittkowski2014} for further examples.

In this paper, we study the dynamics of a one-dimensional ferrogel model far from equilibrium. 
We address questions on the touching and detachment dynamics of magnetic particles.
The dipole-spring model is 
adopted as such a mesoscopic model deals with the configurations of magnetic particles in 
a direct manner: the magnetic particles are explicitly resolved and the distances between them are 
simply related to the lengths of springs attached between them. 
Then a quasi-continuum equation governing the behavior of the system is derived 
based on a term equivalent to the particle density. 
Our main results show that the large-scale chain formation dynamics in the long-time regime 
are governed by shock-wave solutions in the continuum description. 
With the aid of singular perturbation theory~\cite{Witelski1995, Witelski1996} 
in connection with the Stephan problem~\cite{Crank1975, Pego1989}, we can successfully 
quantify the propagation speed. The origin of our regularization and the relation 
to general phase separation dynamics are also discussed.

This paper is organized as follows. In Sec.~\ref{sec:model}, a one-dimensional version of the 
dipole-spring model is introduced. Then we derive a quasi-continuum description of the system 
and discuss its theoretical properties in Sec.~\ref{sec:theory}. Section~\ref{sec:scenario} 
is devoted to illustrate the various observed dynamical scenarios 
and to develop a ``behavioral diagram'' for the different types of dynamics, 
which represents the main result of our study. 
Lastly, a summary and an outlook are given in Sec.~\ref{sec:discussion}.
 
\section{The model} 
\label{sec:model}
\begin{figure}
\includegraphics[width=0.48\textwidth]{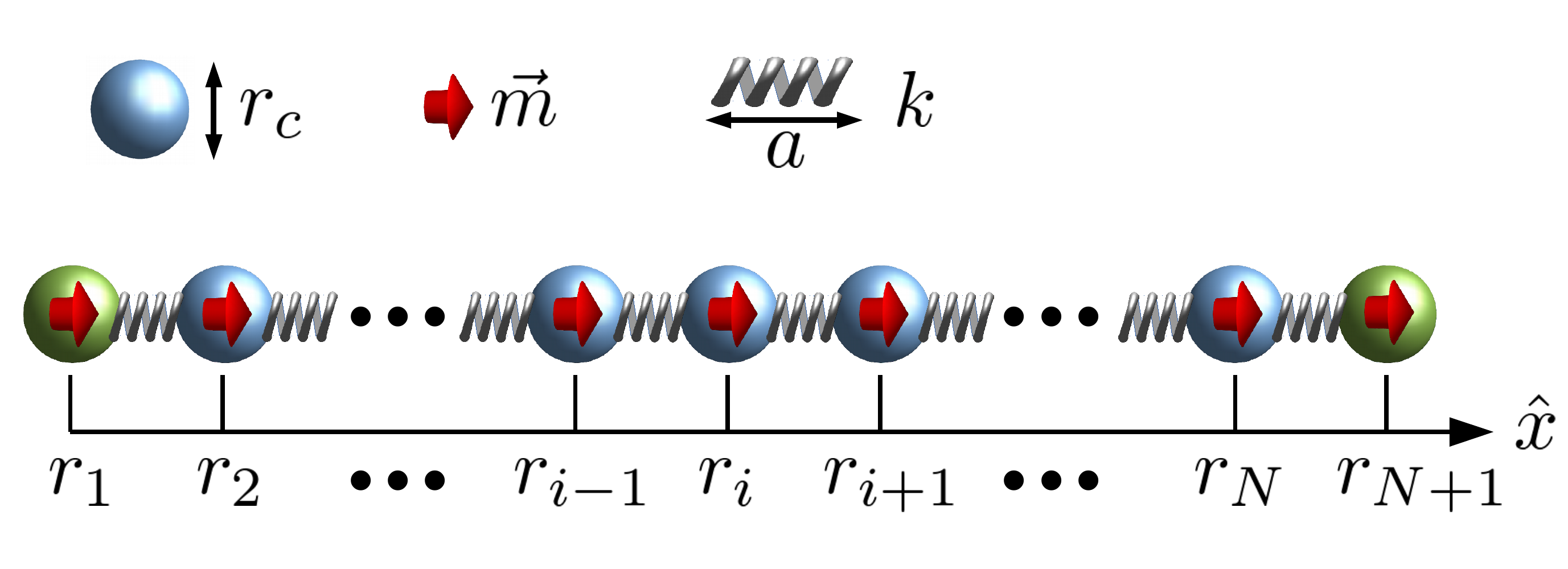}
\caption{\label{fig:model}
A graphical illustration of our model. Blue and green spheres represent the magnetic particles inside and at the boundaries of the system
with the virtual diameter $r_c$, respectively, 
while red arrows correspond to the magnetic dipole moment $\vec{m}$ of the particles. Springs of spring constant $k$ and undeformed length $a$ are
depicted as well. In addition, the convention for quantifying the locations of the particles is also illustrated in this figure.}
\end{figure}

%
Our one-dimensional dipole-spring model for a ferrogel system consists of magnetic particles and 
springs~\cite{Annunziata2013, Pessot2016, Cremer2017}: $N+1$ magnetic particles are connected by 
$N$ harmonic springs, forming a linear straight chain (see Fig.~\ref{fig:model} for a graphical illustration). The number of particles is 
finite so that the chain has definite boundaries at both ends. In this way, we can perturb 
the system by applying forces at the boundaries as in the laboratory. 
The location of the $i$th particle is then represented by $r_i$ for 
$i=1,\ldots, N+1$ and the length of the $i$th spring between the $i$th and the $(i+1)$th 
particles by $r_{i,i+1} \equiv r_{i+1} -r_{i}$ for $i=1, \ldots,N$. The magnetic dipole moment 
$\vec{m}_i$ is assigned to the $i$th particle, which can be any vector in the three-dimensional 
space. Below, after switching to a non-zero value, it will be considered as constant in time.

Following previous studies~\cite{Tarama2014,Pessot2016,Pessot2018}, we consider the overdamped 
dynamics of the dipole-spring model as a function of time $t$, governed by equations of motion 
of the form
\begin{align}
\label{eq:overdamped}
\Gamma \frac{\dd r_i}{\dd t} = - \frac{\partial E_{tot}}{\partial r_i}.
\end{align}
Obviously, the form of the total energy $E_{tot}$ determines the dynamical properties of 
the magnetic chain. In this study, we adopt a simple version of the dipole-spring model in which 
$E_{tot}$ is given by the sum of elastic interactions, magnetic dipole-dipole interactions,
and steric repulsion~\cite{Pessot2016, Cremer2017,Pessot2018}. First, the elastic energy of 
the harmonic springs takes the form of
\begin{align}
\label{eq:e_el}
E_{el}(\{r_{i,i+1}(t)\}) = \frac{k}{2}\sum_{i=1}^N \left( r_{i, i+1}(t) -a \right)^2,
\end{align}
where $k$ is the spring constant and $a$ the length of the springs in the undeformed state.
Second, the magnetic dipole-dipole interaction energy is given as
\begin{align}
\label{eq:e_m}
E_{m}(\{r_{ij} (t)\}) =\frac{\mu_0}{4\pi} &\sum_{j>i} 
 \left[ \frac{(\vec{m}_i \cdot \vec{m}_j)}{\left|\vec{r}_{ij}(t)\right|^3} \right. \nn \\
	 &\left. -\frac{3(\vec{m}_i \cdot \vec{r}_{ij}(t))(\vec{m}_j \cdot \vec{r}_{ij}(t)) }
	{\left|\vec{r}_{ij}(t)\right|^5} \right], 
\end{align}
where $\mu_0$ is the vacuum permeability, $r_{ij}=r_j -r_i$, and 
$\vec{r}_{ij} \equiv r_{ij} \hat{x}$ with the unit vector $\hat{x}$ along the chain axis.
For simplicity, we limit ourselves to the case in which the magnetic moments are identical across 
the whole system, $\vec{m}_i \equiv m (\cos{\theta}\,\hat{x} +\sin{\theta} \cos{\phi}\,\hat{y} 
+\sin{\theta} \sin{\phi}\,\hat{z})$. Then, introducing $b\equiv 1-3\cos^2{\theta}$, 
one can rewrite the magnetic energy in a simpler form
\begin{align}
E_{m}(\{r_{ij} (t)\}) =\frac{\mu_0 b m^2}{4\pi} \sum_{j>i} \frac{1}{\left|r_{ij} (t)\right|^3}. 
\end{align}
Finally, the steric repulsion preventing a collapse of the system under strong magnetic fields,
reads
\begin{align}
\label{eq:e_st}
E_{st}(\{r_{i,i+1}(t)\}) = \sum_{i=1}^N U_{WCA} (r_{i,i+1}(t)),
\end{align}
where $U_{WCA}$ is a modified Weeks-Chandler-Andersen (WCA) type potential
in the form~\cite{Weeks1971, Pessot2016}
\begin{align}
U_{WCA} (r) &= \Theta(r_c-r) \epsilon^s \left[\left( \frac{r}{\sigma^s} \right)^{-12}
	-\left( \frac{r}{\sigma^s} \right)^{-6} \right. \nn \\
	&\left. -\left( \frac{r_c}{\sigma^s} \right)^{-12} +\left( \frac{r_c}{\sigma^s} \right)^{-6} 
	-\frac{c^s (r-r_c)^2}{2} \right],
\end{align}
with the Heaviside step function $\Theta$ and a cutoff distance $r_c$. Here, $\sigma_s$ and 
$c_s$ are chosen such that $U'_{WCA} (r_c)=0$ and $U''_{WCA}(r_c)=0$~\cite{Pessot2016}. 
The parameter $\epsilon^s$ characterizes the strength of the steric repulsion. Now, one can find 
a set of dynamic equations, substituting the above definitions directly into 
Eq.~\eqref{eq:overdamped}. Those equations for particles inside as well as at the boundaries 
of the system are described in detail in Appendix~\ref{ap:eq_motion}.

\section{Formulation of a quasi-continuum theory}
\label{sec:theory}
%
To be able to develop a continuum description of the system, we as a major simplification
cut the long-range magnetic interaction beyond the nearest-neighbor interaction. 
We have confirmed from particle-resolved simulations that the overall dynamics with the 
nearest-neighbor and long-range magnetic interactions are qualitatively equivalent to each other,
as far as uniform configurations are adopted as initial conditions
(see Sec.~\ref{sec:init_bound}).

\subsection{Continuum equation}
\label{sec:continuum}
%
Now the energy is only a function of the distances between adjacent particles as follows: 
\begin{align}
E_{tot} (\{r_{i, i+1} (t)\}) = \sum_{i=1}^{N} e\, (r_{i,i+1} (t)),
\end{align}
where $e$ is the pairwise energy given by
\begin{align} \label{eq:energy_density}
e(r)=\frac{k}{2}(r-a)^2 +\frac{\mu_0 bm^2}{4\pi}\frac{1}{r^3}+U_{WCA}(r).
\end{align}
The direct analysis of this pairwise energy landscape will help 
in understanding the equilibrium states as well as the dynamics of the systems 
and therefore constitutes one of the essential parts of the continuum theory.
We address this issue in detail in Sec.~\ref{sec:theory3}.

We then seek a continuum description of the system, introducing a continuous variable $x$, 
a positional field $r(x,t)$, and its derivatives with respect to $x$, i.e., $r_x$, $r_{xx}$, 
and so on. Following a standard transformation rule
$\displaystyle \sum_{i=1}^{N+1} \to \int_0^{N} \dd x$, $r_{i+1}-r_{i} \to \partial r/\partial x$, 
and $\partial/\partial r_i \to \delta/\delta r$ (see, e.g., Ref.~\onlinecite{Doi1986}),
we directly obtain from Eq.~\eqref{eq:overdamped} a fully continuum equation
\begin{align}
\label{eq:cont_typical}
\Gamma r_t(x,t) = {e}^{(2)}(r_x(x,t)) r_{xx}(x,t),
\end{align}
where $e^{(i)}(r)\equiv \left( \frac{\partial}{\partial r}\right)^i e(r)$ for a general natural number $j$.
If the explicit form of the energy is inserted, the continuum equation reads
\begin{align} \label{eq:cont_DS_leading}
\Gamma r_t =& k r_{xx} +  \frac{3 \mu_0 b m^2}{\pi} \frac{r_{xx}}{(r_x )^5} 
	+ \epsilon^s r_{xx} \Theta (r_c-r_x) \nn \\ 
	&\times \left[ 
		\frac{156}{(\sigma^s)^2} \left( \frac{r_x}{\sigma^s} \right)^{-14} 
	 -\frac{42}{(\sigma^s)^2} \left( \frac{r_x}{\sigma^s} \right)^{-8} -c^s
	\right].
\end{align}
Two important characteristics of the continuum equation are summarized as follows: First 
Eq.~\eqref{eq:cont_typical} takes the form of a diffusion equation. However, the diffusion 
coefficient $e^{(2)}(r_x (x,t))$ may have a negative value depending on the value of $m$. 
Second, the variable $r_x$, which determines the sign of $e^{(2)}$, is closely connected 
to the particle density via the relation $\rho(x,t)=1/r_x (x,t)$. Therefore, the particle density 
controls the dynamics. 

\subsection{Regularization}
\label{sec:regularization}
%
We note that, if there exists a range with $e^{(2)} (r_x) <0$, the continuum equation becomes 
a type of the forward-backward heat equation which does not necessarily have a unique 
solution~\cite{Hollig1983}. It is then mandatory to include an additional term as 
a regularization, which should be specific for each given physical problem~\cite{Barenblatt1993}. 
In our case, the regularization stems from the discrete nature of the system, 
similarly to the lattice regularization in critical phenomena. 

Indeed, the transformation rule $r_{i+1}-r_{i} \to \partial r/\partial x$ involves
a truncation of higher order terms $r_{xx}$, $r_{xxx}$, and so on, neglecting 
corrections from the discreteness of the system. Here, we explicitly take such corrections
into account. We consider the differences $\Delta_i r \equiv r_{i+1}-r_{i}$ instead of 
the differential $\partial r/\partial x$ and utilize the functional-derivative technique 
for discrete variables. As expected, this approach leads to the equations of motion 
for $i=2,\ldots,N$, described in Appendix~\ref{ap:eq_motion}, which formally read
\begin{align}
\Gamma \frac{\dd r_i}{\dd t} = \Delta_{i-1} e^{(1)} (r_{i,i+1} (t)).
\end{align}

Now we probe a continuum description via a transformation from the discrete variable 
$i=1, 2, \ldots, N+1$ to a continuous variable $x$ defined in a domain $0 < x < x_{\rm max}$. 
We choose the midpoint rule to weight equivalently the forces from the right 
and left sides of the particle under consideration. In other words, a point $i$ in the discrete 
description corresponds to a range $[x-\Delta x/2, x+\Delta x/2]$ in the continuum description,
where $\Delta x$ controls the discreteness of the system. 
If an asymmetric rule is considered, particles may prefer a motion in a certain direction.
Then, with a transformation of the form $x\equiv (i-\frac{1}{2})\Delta x$,
the domain on which the newly introduced continuous variable $x$ is defined
is determined as $0\leq x \leq x_{\rm max}$ with $x_{\rm max} \equiv (N+1)\Delta x$. 
Here, we take $x_{\rm max} = 1$ by further setting $\Delta x \equiv 1/(N+1)$. 
In this way, a continuum limit is achieved via $\Delta x \to 0$ or $N\to \infty$. 
In contrast to that, $r_{xx}$ and higher order derivatives were neglected in the case of the 
previous transformation rule to the domain $0 \leq x \leq N$ with $\Delta x =1$ 
in Sec.~\ref{sec:continuum}, which was used to derive Eq.~\eqref{eq:cont_typical}.

We set up a transformation rule in the form $r_i (t) \to r(x,t)/(\Delta x)$,
relating the length scale of springs to the corresponding continuous variable $r_x$ as
$r_{i,i+1} (t) = r_{i+1}(t)-r_{i}(t) \to \frac{r(x+\Delta x,t)-r(x,t)}{\Delta x} 
=r_x (x,t) +\frac{\Delta x}{2}r_{xx} (x,t)+\cdots$.
Then we obtain a continuum description from the discrete form in Eq.~\eqref{eq:cont_typical} 
as follows:
\begin{align} \label{eq:disc_form_det}
{\Gamma r_t}=(\Delta x)
	&\left\{ e^{(1)}\left( \frac{r(x+\Delta x,t)-r(x,t)}{\Delta x} \right) \right. \nn \\
	&\left. -e^{(1)}\left( \frac{r(x,t)-r(x-\Delta x,t)}{\Delta x} \right) \right\}.
\end{align}
Rescaling $x \to x/(\Delta x)$ from the domain $0 \leq x \leq1$ to $0\leq x \leq N+1$, 
one indeed recovers to leading order the continuum description of the dynamics described by 
Eq.~\eqref{eq:cont_typical}. This can be easily checked by expanding Eq.~\eqref{eq:disc_form_det}
in terms of $\Delta x$ as
\begin{widetext}
\begin{align} \label{eq:cont_series}
\Gamma r_t =&-(\Delta x)\sum_{i=1}^\infty \frac{(-1)^i}{i!}
	\left\{ \frac{r(x+\Delta x,t) -2r(x,t)+r(x-\Delta x,t)}{\Delta x} \right\}^i 
	e^{(i+1)} \left( \frac{r(x+\Delta x,t) -r(x,t)}{\Delta x} \right) \nn \\
	=&-(\Delta x)\sum_{i=1}^\infty \frac{(-1)^i}{i!}
		\left\{ 2\sum_{j=1}^\infty \frac{(\Delta x)^{2j-1}}{(2j)!}r^{(2j)}(x,t)	\right\}^i 
		 \left\{ \sum_{k=0}^\infty \frac{e^{(i+1+k)} \left( r_x (x,t) \right)}{k!}
		\left( \sum_{l=2}^\infty \frac{(\Delta x)^{(l-1)}}{l!}r^{(l)}(x,t) \right)^k \right\},
\end{align}
\end{widetext}
where $r^{(i)} \equiv \left( \frac{\partial}{\partial x}\right)^i r$.
In this way, we maintain aspects of the discrete nature of the system in a quasi-continuum 
description by letting $\Delta x$ become small but finite. Accordingly,
we obtain regularization terms to Eq.~\eqref{eq:cont_typical} from the second- and even 
higher-order contributions in Eq.~\eqref{eq:cont_series}. 

Now one could seek for a precise solution, including all terms in Eq.~\eqref{eq:cont_series}. 
Instead, one may truncate them at a certain order, searching for approximate solutions. 
At this point, we encounter the mathematical issue that different 
regularization forms may differently alter the dynamics of the forward-backward heat 
equation~\cite{Novick-Cohen1991, Barenblatt1993}. Rather than rigorously investigating 
this issue, we here take a pragmatic way using the next-order correction 
as a feasible form of regularization. This leads to a regularized equation
\begin{align} \label{eq:regularization}
\Gamma r_t &\approx (\Delta x)^2 e^{(2)} \left(r_x \right)\, r_{xx} 
	+ (\Delta x)^4 \left[ \frac{1}{12}e^{(2)} (r_x)\,r_{xxxx} \right. \nn \\ 
	&\left. +\frac{1}{6}e^{(3)} (r_x)\,r_{xx} r_{xxx} 
	+\frac{1}{24}e^{(4)} (r_x)\,(r_{xx})^3 \right].
\end{align}
In the end, a certain type of regularization is necessary to evaluate the equations. Our approach
has the strong benefit of being fully systematic.

\subsection{Initial and boundary conditions}
\label{sec:init_bound}
%
For simplicity, we only consider uniform initial configurations, i.e., $r_x(x,t=0)=v^{\rm init}$ 
with constant $v^{\rm init}$.
Boundary conditions in the quasi-continuum theory are carefully determined from 
the model as follows. First, it is clear that the quasi-continuum equation, e.g., 
Eq.~\eqref{eq:cont_series}, applies to the interior particles. For the boundary particles,
additional rules are necessary. Specifically, the boundary particles ($i=1, N+1$) at the 
left/right ends are governed by
\begin{align}\label{eq:ptles_boundary}
\Gamma \frac{{\rm d}r_1}{{\rm d}t}=\frac{\partial e(r_{1,2} (t))}{\partial r_2}, \quad
\Gamma \frac{{\rm d}r_{N+1}}{{\rm d}t}=-\frac{\partial e(r_{N,N+1} (t))}{\partial r_{N+1}},
\end{align}
while we have
\begin{align}\label{eq:ptles_inside}
\Gamma \frac{{\rm d}r_i}{{\rm d}t}=
	-\frac{\partial e(r_{i-1,i} (t))}{\partial r_i}-\frac{\partial e(r_{i,i+1} (t))}{\partial r_i},
\end{align}
for the particles inside the chain. To fill this gap and make the quasi-continuum equations of 
motion applicable to the boundary particles as well, we introduce hypothetical particles 
$i=0, N+2$, following the procedure in Ref.~\onlinecite{Doi1986}. 
Indeed, Eq.~\eqref{eq:ptles_boundary} takes the same form as Eq.~\eqref{eq:ptles_inside}, 
if the location of the hypothetical particles, 
$r_0$ and $r_{N+2}$, are given by the positions satisfying
\begin{align}
\frac{\partial e(r_{0,1} (t))}{\partial r_1}=0, \quad 
\frac{\partial e(r_{N+1,N+2} (t))}{\partial r_{N+1}} =0,
\end{align}
so that the forces from the hypothetical to the boundary particles are zero.
In the continuum limit, the above relations to lowest order take the form
\begin{align}
\label{eq:boundary_condition}
\left. e^{(1)}(r_x) \right|_{x =\delta \Omega} = 0,
\end{align}
where $\delta \Omega =\{ 0, 1 \}$.

In practice, the dynamics together with the initial and boundary conditions described above 
could be interpreted in two different ways as follows: First, one may imagine an infinitely long chain 
at an unstable fixed point. In this case, the dynamics are initiated by \emph{cutting} off 
the outer parts of the chain at the boundaries. Secondly, a stable finite chain 
with definite boundaries may be considered from the beginning. 
With $m=0$, for instance, we attain a homogeneous chain as the equilibrium configuration,
in which the distances between adjacent particles are equivalent 
to the equilibrium spring lengths $a$. The dynamics of the system is then initiated 
by turning on an external magnetic field. This accords with a \emph{quenching} procedure.
In both cases, the interior particles are still in the state of the unstable fixed point
directly after the initiation procedures, because the forces from the left and the right 
particles are well balanced due to the initial homogeneity. This is true as long as only 
nearest-neighbor magnetic interactions are taken into account. The long-range magnetic 
interactions slightly affect this picture. However, also in our test simulations including
long-range magnetic interactions, we have not observed qualitative differences.
As far as the uniform initial configurations are considered, 
it is always the boundaries from which the dynamics are initiated: 
for the particles at the left/right boundaries, forces are only acted from the particle on 
the right/left side at the moment of cutting or quenching.

\subsection{Pairwise energy landscape}
\label{sec:theory3}
\begin{figure}
\includegraphics[width=0.48\textwidth]{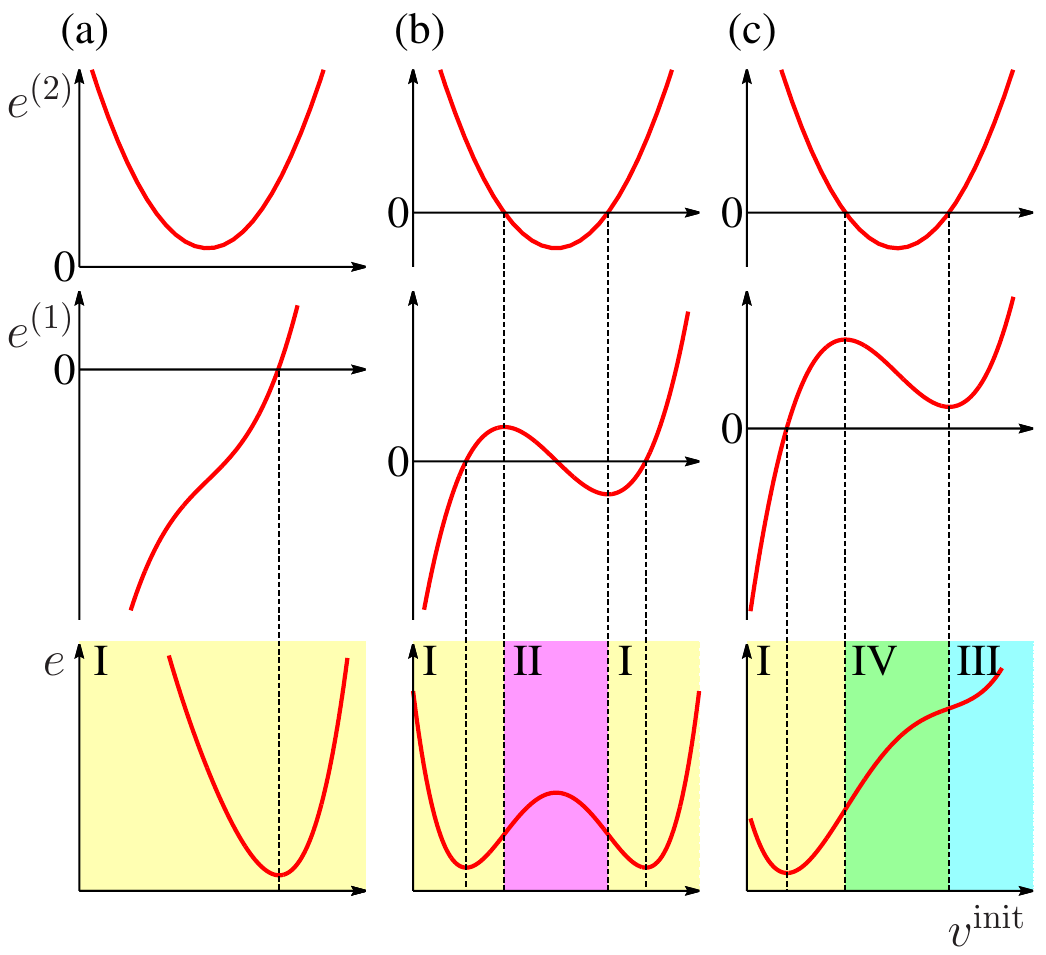}
\caption{\label{fig:landscape}
Schematic illustration of the qualitatively different landscapes of the pairwise energies, 
leading to qualitatively different initial and long-time dynamical behaviors. 
While regions of $e^{(2)}(v^{\rm init}) <0$
in the first row represent spinodal-like intervals, points at $e^{(1)}(v^{\rm init})=0$ 
in the second row indicate local extrema in the energy density 
landscapes (see $e$ in the third row). Based on the number of minimum points and the existence
of the spinodal-like interval, the landscapes are classified into three qualitative categories,
depicted in (a), (b), and (c). Further analyzing the sign of $e^{(2)}$ 
and the location of the minimum points, we classify the behavior 
according to the value of $v^{\rm init}$ into different regimes I, II, III, and IV shaded by
different colors. See the text for the dynamic scenario corresponding to each regime.}
\end{figure}

%
Mathematically, the pairwise energy $e$ plays a similar role as the free energy does in 
thermodynamics as thermal fluctuations of the particles~\cite{Cremer2017} are neglected
in the present study. Above all, (mechanical) equilibrium configurations correspond to the minimum 
points in the landscape of the pairwise energy, which are determined to lowest order from the 
condition $e^{(1)}(r_x)=0$ in the continuum limit. We note that uniform equilibrium 
configurations automatically satisfy the static equilibrium condition $r_t =0$ of our 
regularized equation, Eq.~\eqref{eq:regularization}.
In this study, we are considering a double-well potential, 
with at least one and at most two locally stable fixed (equilibrium) points: 
one corresponds to the configuration in which the particles touch each other (high density) 
while the particles stay away from each other (low density) in the other configuration. 
Tuning the magnetic moment $m$, one can modulate 
the number of stable equilibrium points~\cite{Annunziata2013}. 
Furthermore, the diffusion coefficient $e^{(2)}(r_x)$, 
the sign of which plays a significant role as already discussed in Sec.~\ref{sec:regularization}, 
is also modulated by the values of $m$. Together with the trivial one for the initial condition, 
we take into account two independent control parameters, namely, 
the magnetic moment $m$ and the initial uniform distance between adjacent 
particle pairs $v^{\rm init}$. Keeping these in mind, we classify 
the landscapes of the pairwise energy into three categories as shown in Fig.~\ref{fig:landscape}.

First, we confirm that $e^{(2)}(v^{\rm init}) >0$ for all the $v^{\rm init}$ values, if the 
magnetic moment $m$ is very small [Fig.~\ref{fig:landscape}(a)]. In this case, there exists 
only one minimum in the pairwise energy landscape and the whole range of $v^{\rm init}$ (colored 
in yellow) belongs to the basin of attraction of the minimum point. From now on, the term 
Scenario I is used to indicate relaxation dynamics to the stable equilibrium corresponding 
to this case.

If the magnetic moment is very strong [Fig.~\ref{fig:landscape}(c)], once again there is only one 
stable fixed point which corresponds to a hardened touching configuration of the 
particles~\cite{Annunziata2013}. In this case, however, there 
exists a range with $e^{(2)}(v^{\rm init}) <0$ (shaded in green) analogous to the spinodal 
interval, which divides the range of $v^{\rm init}$ with $e^{(2)}(v_{\rm init}) > 0$ 
into two regions: a high-density region 
(in yellow) forming a basin of the only minimum point and a low-density one (in cyan) 
separated from the fixed point. Among these three regions, the dynamics around 
the equilibrium (yellow) is equivalent to Scenario I, while the dynamics starting 
from the spinodal-like interval (green) and the low-density regime (cyan) are qualitatively 
different from Scenario I and, respectively, referred to as Scenario III and IV in this paper. 

If we consider magnetic moments lower than for the strong-$m$ regime,
bistable landscapes appear [Fig.~\ref{fig:landscape}(b)]. Once again, separated regions 
with positive diffusion coefficients (in yellow as before) correspond to Scenario I. 
In contrast to that, the dynamics starting from the spinodal-like interval in between 
(shaded in magenta) exhibits a new behavior which is called Scenario II henceforth. 
Between the bistable [Fig.~\ref{fig:landscape}(b)] and very-weak-$m$ regime 
[Fig.~\ref{fig:landscape}(a)], there is an interval with an energy landscape 
similar to an inversion (e.g., by $v^{\rm init} \to 1/v^{\rm init}$) 
of the abscissa in Fig.~\ref{fig:landscape}(c). As one may expect, no further dynamics
qualitatively different from the ones of Scenario I, III, and IV are observed in this case.

\section{Scenarios}
\label{sec:scenario}
%
Using Eq.~\eqref{eq:cont_DS_leading}, its regularized quasi-continuum version 
Eq.~\eqref{eq:regularization}, and the analysis of the pairwise energy landscapes 
in Sec.~\ref{sec:theory3}, we now describe the dynamical scenarios in detail. 
Even though the quasi-continuum theory is developed using the variable $x$, simulation 
results are presented in terms of the density as a function of the location of the particles 
in real space, namely, $\rho_i\equiv 1/r_{i,i+1}$ versus $r_i$, if not specified otherwise. 
Similarly, we also use the terms $\rho(x)\equiv 1/r_x$, $\rho^{\rm init} \equiv 1/v^{\rm init}$
to lowest order, and so on.
Henceforth, time, length, and energies are rendered dimensionless setting $\Gamma/k$, $r_c$, 
and $kr_c^2$ as units of measurement, respectively. In this unit system, a density of 
$\rho/\rho_c =1$ with $\rho_c \equiv 1/r_c$ indicates the set of touching adjacent magnetic 
particles. Moreover, magnetic moments are then measured in a unit of 
$m_0 \equiv \sqrt{4\pi k r_c^5 /\mu_0}$. In plotting the figures for 
additional particle-resolved simulation results, values of $a/r_c=2.5$, $\epsilon^s/(kr_c^2) =1$, and $b=-2$ 
(assuming that the dipole moments are parallel to the chain axis and all pointing into the 
same direction) have been used and red cross symbols 
in the figures represent initial density distributions. Even though only results for $N=100$ 
are shown, we have observed equivalent dynamics simulating systems with $N=$200, 400, and 800.

\subsection{Scenario I: simple relaxation}
\label{sec:scen1}
\begin{figure}
\includegraphics[width=0.48\textwidth]{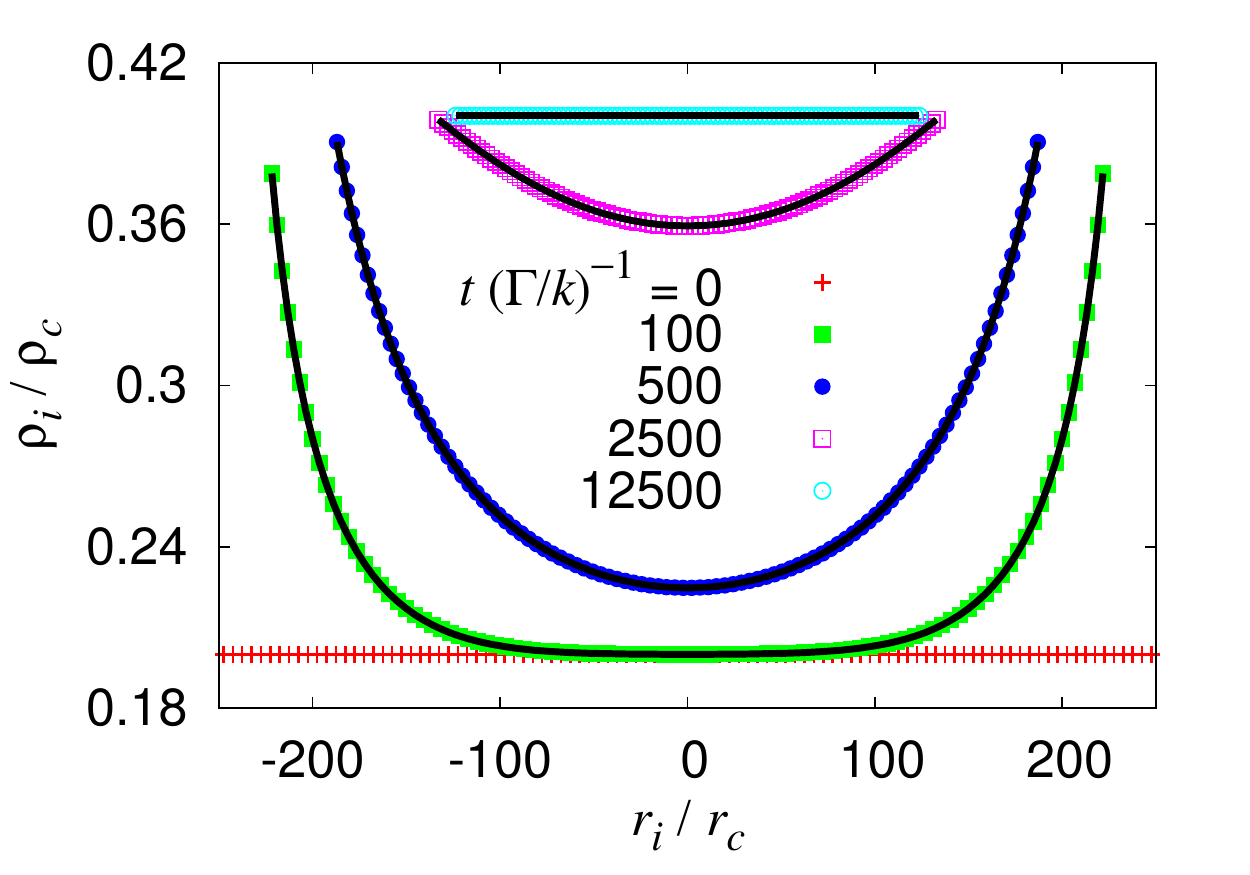}
\caption{\label{fig:scen1}
Time evolution of density profile for Scenario I. Density distributions extracted from 
the particle-resolved simulations are presented by symbols while black lines show the 
numerical solution of Eq.~\eqref{eq:cont_DS_leading}. 
Here, $m=0.1 m_0$ and $\rho_i^{\rm init}=0.2 \rho_c$. 
Agreement between particle simulation and numerical solution of the theory is manifested clearly. 
These results are also depicted in more detail in MOVIE I of ESI~\cite{supple}.}
\end{figure}

%
We first describe the scenario in the yellow regimes in Fig.~\ref{fig:landscape},
in which the uniform initial configuration belongs to the basin 
of attraction of the equilibrium point given by $e^{(1)}(\rho_{eq})=0$. Moreover,
$e^{(2)}(\rho)$ is always positive during the whole time evolution and the regularization 
is not necessary: 
direct integration of Eq.~\eqref{eq:cont_DS_leading} yields very good agreement with  
particle-resolved simulations as shown in Fig.~\ref{fig:scen1}.
The density profiles evolving in time can be either concave ($\rho^{\rm init} > \rho_{eq}$)
or convex ($\rho^{\rm init} < \rho_{eq}$).

\subsection{Scenario II: pair formation}
\label{sec:scen2}
\begin{figure}
\includegraphics[width=0.48\textwidth]{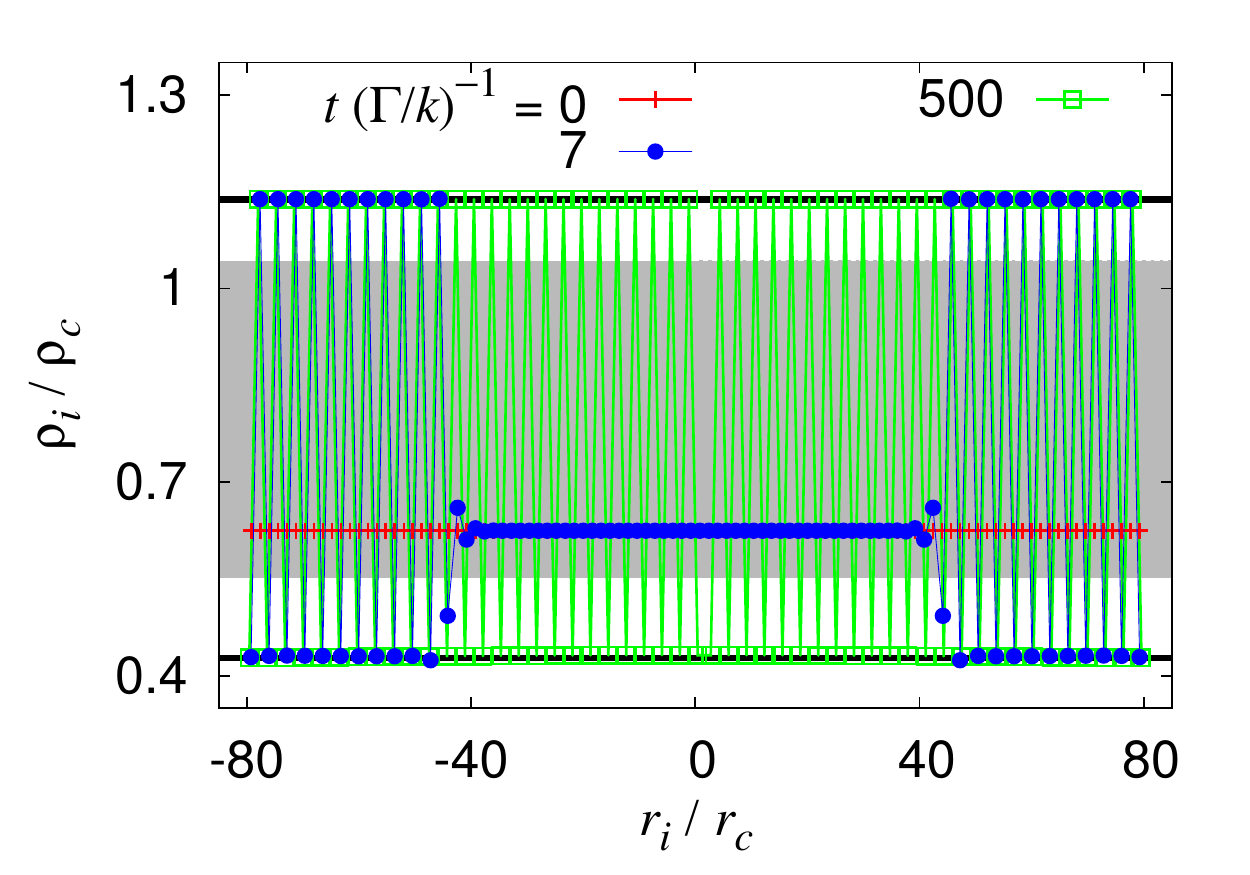}
\caption{\label{fig:scen2}
Density distributions in Scenario II. Symbols display the simulation results and thick black lines 
represent the uniform configurations corresponding to the local minimum points of the 
pairwise energy. The region shaded in gray represents the spinodal-like interval. In this figure, 
$m=0.9 m_0$ and $\rho_i^{\rm init}=0.625 \rho_c$. For more details, see MOVIE II 
of ESI~\cite{supple}.}
\end{figure}

%
Now, we consider the dynamics in the bistable regime with initial configurations in the 
spinodal-like interval, i.e., the range satisfying $e^{(2)}(\rho^{\rm init}) < 0$. 
As depicted by black lines in Fig.~\ref{fig:scen2}, there are two different states 
of energetic minima, in both of which the corresponding configurations are uniform. 
In the particle-resolved simulations, we observe formations of particle pairs.
As represented by high-density points in Fig.~\ref{fig:scen2}, the pairs consist of two 
touching particles, appearing in a row along the chain. Both of the two densities computed 
from the pairs as well as from the stretched springs between the pairs coincide with the 
density values of the minimum points in the pairwise energy.
This indicates that the heterogeneous configurations are stable in the discrete systems
in the absence of fluctuations. Therefore, in the particle-resolved simulations, 
the relaxation to the global minimum state with a uniform configuration is not observed.
In addition, we note that, near the spinodal lines of $e^{(2)}=0$, clusters with a number of 
touching particles larger than 2 (high-density spinodal line) or stretched springs with only one magnetic particle in between (low-density spinodal line) are observed as stable configurations in the simulations.

We then turn to the continuum theory. For this scenario, a regularization is mandatory. 
There are two different candidates for the boundary condition, both of which 
satisfy Eq.~\eqref{eq:boundary_condition}, as we consider the bistable regime. 
Here, let us take the global minimum state as a boundary condition.
Then it is observed that numerical solutions of the continuum equation
(see Appendix~\ref{ap:numerics} for further details) converge to the global 
minimum state with a uniform configuration in contrast to the particle-resolved simulations.
Such a disagreement may imply a failure of the continuum theory in providing a full description 
in this regime. Indeed, it is well known from the $\Gamma$-convergence theory that the solutions 
to the Cahn-Hilliard equation asymptotically approach to the global minimum 
point~\cite{Modica1987, Pego1989, Braides2002}. Similarly, we conjecture that the asymptotic 
solutions to our quasi-continuum theory are given as the uniform configuration at the global 
minimum point. This may, for instance, be due to our non-exact regularization terms in 
the quasi-continuum description together with the numerical scheme adopted 
in the integration of the continuum equation that includes additional diffusion.
Thus, the bistability is at present only visible in our discrete particle simulations.

For further insight, we inspect the individual particle dynamics. Let us consider 
a particle and its two nearest neighbors as well as the two springs connecting them.
With the two distances between the two particle pairs, $r_1$ and $r_2$, 
the corresponding energy can be written as $E_{in}(r_1,r_2) = e(r_1)+e(r_2)$. 
Then introducing new variables $L\equiv r_1+r_2$ and $l=r_1-r_2$, we first confirm that 
the state of $l=0$ with a homogeneous configuration corresponds to a fixed point 
of the dynamics because
\begin{align}
\left. \frac{\partial E_{in}}{\partial l}\right|_{l=0} = \left[ \frac{1}{2}e^{(1)}\left( \frac{L+l}{2}\right)
- \frac{1}{2}e^{(1)}\left( \frac{L-l}{2}\right) \right]_{l=0} = 0.
\end{align}
Meanwhile, the fixed point of $l=0$ is unstable if $e^{(2)}(L/2)=e^{(2)}(r_1)=e^{(2)}(r_2) <0$ 
as one can easily verify from the corresponding Hessian matrix
\begin{align}
\left( {\begin{array}{cc} \frac{1}{2}e^{(2)}(L/2) & 0 \\
	  0  & \frac{1}{2}e^{(2)}(L/2) \\  \end{array} } \right).
\end{align}
For $e^{(2)}(L/2)<0$, as in this case, it is straightforward to describe the onset of 
the scenario: dynamics initiated from the 
boundary (as already discussed in Sec.~\ref{sec:init_bound}) penetrates into the inner part 
of the chain, perturbing particles in the local maximum state. 
Then one may expect a heterogeneity in the configuration (i.e., $l\neq 0$) 
as a consequence of the above spinodal-like decomposition mechanism, 
which underlies the formation of touching particle pairs. As shown in Fig.~\ref{fig:scen2}, 
densities for touching pairs and for the stretched springs between pairs agree well with the
values of the two local minimum points. Consequently, the resulting configuration remains stable 
once the localized spinodal-like decomposition dynamics are accomplished.


\subsection{Scenario III: shock-wave propagation}
\label{sec:scenario3}
\begin{figure*}
\includegraphics[width=0.48\textwidth]{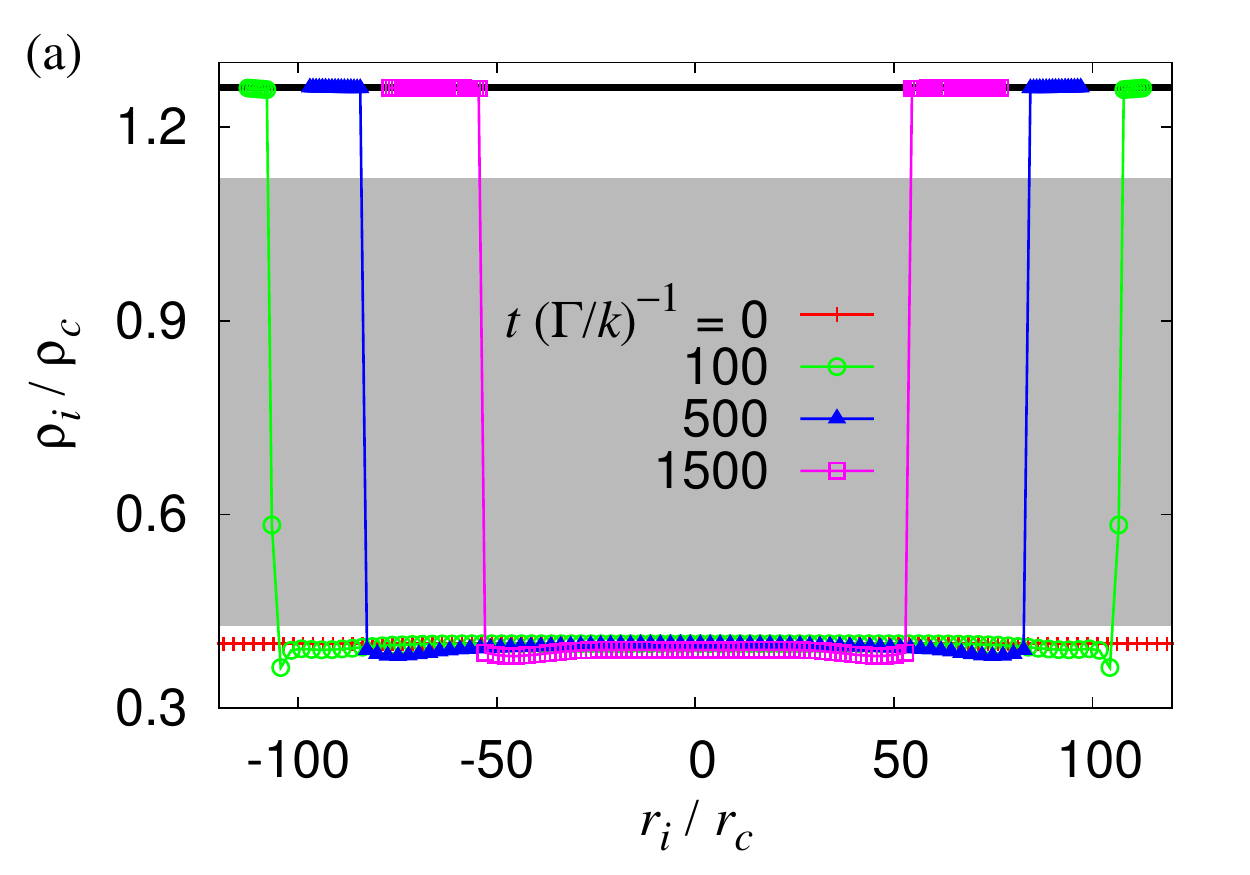}
\includegraphics[width=0.48\textwidth]{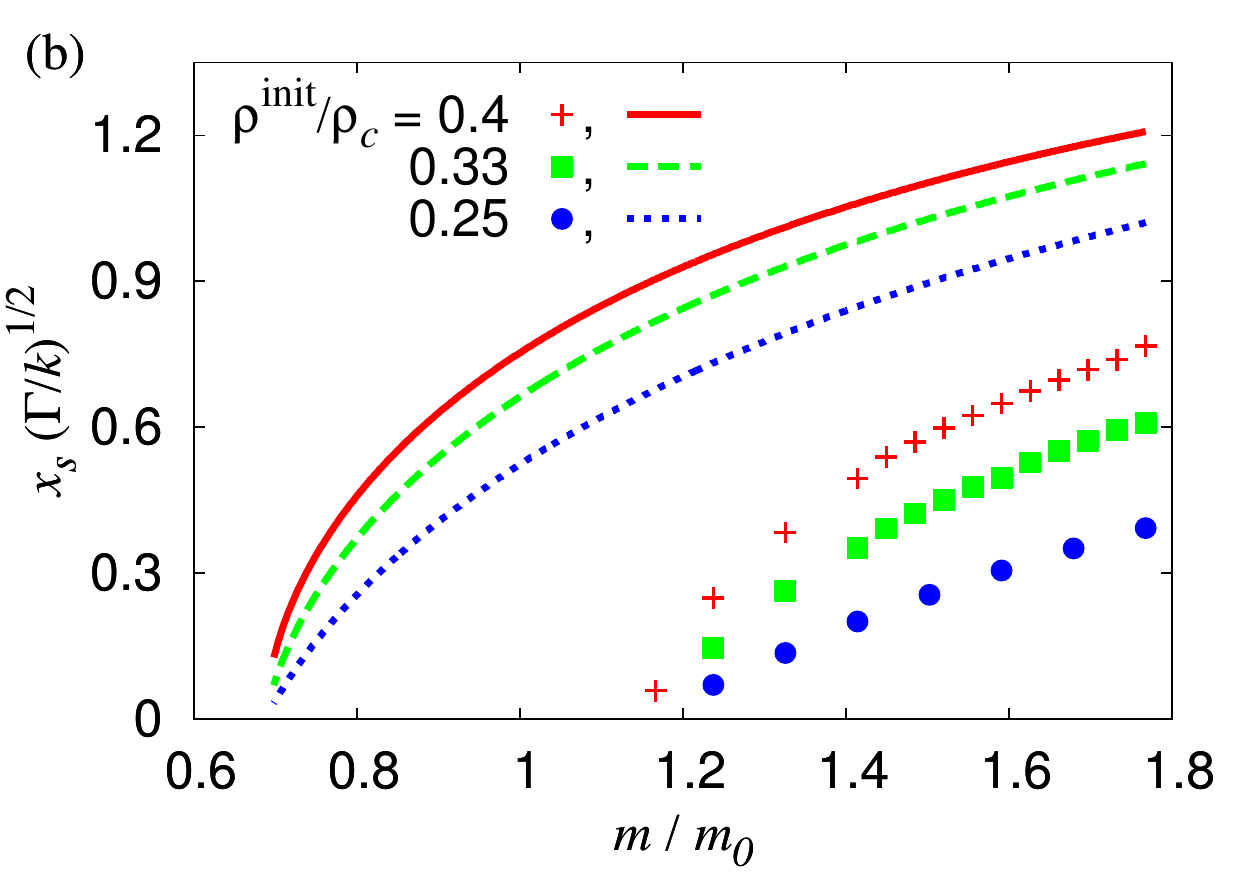}
\caption{\label{fig:scen3}
Results for Scenario III of shock-wave propagation. (a) Density distributions for $m=1.7 m_0$ and 
$\rho_i^{\rm init}=0.4 \rho_c$ are shown, indicating high-density clusters at the ends and 
a low-density region on the inside. Symbols represent the particle-resolved simulation results 
while the equilibrium density computed from the theory is represented by a black line. 
Sharp interfaces between high- and low-density clusters are manifested clearly. 
A movie (MOVIE III) describing the time evolution can be found in ESI~\cite{supple}. (b) Values of 
the coefficient $x_s$, obtained from the particle-resolved simulations (symbols) 
and the singular perturbation theory (lines), are compared to each other.}
\end{figure*}

%
In this scenario, the most important feature observed in the particle-resolved 
simulations is the generation of sharp interfaces which divide the chain into macroscopic 
high-density clusters and stretched low-density configurations, as shown in Fig.~\ref{fig:scen3}. 
Specifically, we observe movements of the interfaces between these regions, which are 
initially formed at the ends of the chain. Such movements or propagations of the interfaces, 
for instance, in the regime of strong magnetic fields [cyan in Fig.~\ref{fig:landscape}(c)], 
make the high-density clusters of touching particles grow towards the center of the chain. 
As one can see, the widths of interfaces are of the order of the distance between adjacent 
particles. Before we proceed, we note that, in this section as well as in Sec.~\ref{sec:scenario4}
devoted to Scenario IV, only the touching dynamics are analyzed. The extension of the discussion
to the detaching dynamics corresponding to the case between very weak or vanishing $m$ and the 
intermediate bistable regime would be straightforward.

According to the analysis on the level of individual particles, the dynamics are initiated 
from the boundary as before. In contrast to Scenario II, however, the perturbation from the 
boundary does not affect the particles inside immediately as they roughly remain 
in a locally stable state. If the effects from the boundaries are not too strong 
(this corresponds to Scenario I, in which the initial state already belongs to 
the basin of the equilibrium point), then the relation $e^{(2)}(L/2) > 0$ can still be satisfied, 
keeping the configuration somewhat uniform. As a consequence, in Scenario I particles persistently 
move towards the center during the whole dynamics.

As a new feature, in Scenario III, the distortion at the boundaries is strong enough due to such
a large difference between the initial and equilibrium densities that the stability of the 
uniform configuration can be disturbed. In this case, the $l=0$ (homogeneous) configuration 
becomes unstable for particles at interfaces. With this mechanism, 
the particles at interfaces can move into the direction opposite to the motion of most other 
particles in this half of the chain as well as of the interface, resulting in touching to the 
high-density cluster at the corresponding end of the chain. Subsequently, a sharp undershoot is 
developed in the density distribution at the interfaces.

In terms of the continuum theory, this scenario corresponds to shock-wave
propagations~\cite{Witelski1995}. With a specific regularization, we are able to 
describe the shock with the aid of singular perturbation theory. Here, we briefly summarize 
the procedure (see Appendix~\ref{ap:singular} for the details). According to singular perturbation 
theory, the structure of the shock is quantified by the values of $r_x$ behind and in front of 
the discontinuity or $v_-$ and $v_+$ as defined in Appendix~\ref{ap:singular}, 
which should satisfy $e^{(1)} (v_-)=e^{(1)} (v_+)$.
Among the candidates satisfying the condition, certain values of $e^{(1)}(v_\pm)$ are selected,
depending on the specific form of regularization. 
For the regularization in Eq.~\eqref{eq:regularization}, 
we find that the shock wave satisfies the equal area rule~\cite{Pego1989} or 
equivalently the common tangent construction (see, e.g., Refs.~\onlinecite{Witelski1996} and~\onlinecite{Wittkowski2014} 
for other types of solution). From the determined values of $v_\pm$, we can then compute the 
similarity coefficient $x_s$ which means the factor in a similarity relation of the type
$N_T =x_s\sqrt{t}$, where $N_T$ is the number of the particles in the high-density cluster. 
As this coefficient determines how fast the shock-wave propagates, 
it is of interest to probe quantitatively its values which are presented
in Fig.~\ref{fig:scen3} (b). As one can see, the overall 
behavior is described qualitatively by the theory, but with non-negligible errors. Regarding the 
fact that here we consider the dynamics near a singularity, this type of error seems
to be acceptable.

In addition to that, we can further classify the density profiles of this scenario into two cases:
The first one corresponds to the case of $\rho^{\rm init} < \rho_+\equiv 1/v_+$ 
to lowest order. As the outer layer solution should connect the initial condition 
$v^{\rm init}$ and $v_+$, 
the existence of an undershoot in the density profile at the shock is expected. 
Considering the conservation of particles involved in determining the shock 
structure~\cite{Crank1975,Pego1989,Witelski1995}, we speculate 
that a mechanism similar to the generation of depletion regions in solidification 
processes ahead of the solidification front~\cite{Sandomirski2011} seems to play a role 
in this undershoot generation. If the initial density is high enough, such an undershoot
disappears and the solution becomes monotonic. In particle-resolved simulations, 
one may take the concavity/convexity of the interior part of chains as an index to identify 
the existence of the undershoot in density profiles.

\subsection{Scenario IV: shock wave of pairs}
\label{sec:scenario4}
\begin{figure}
\includegraphics[width=0.48\textwidth]{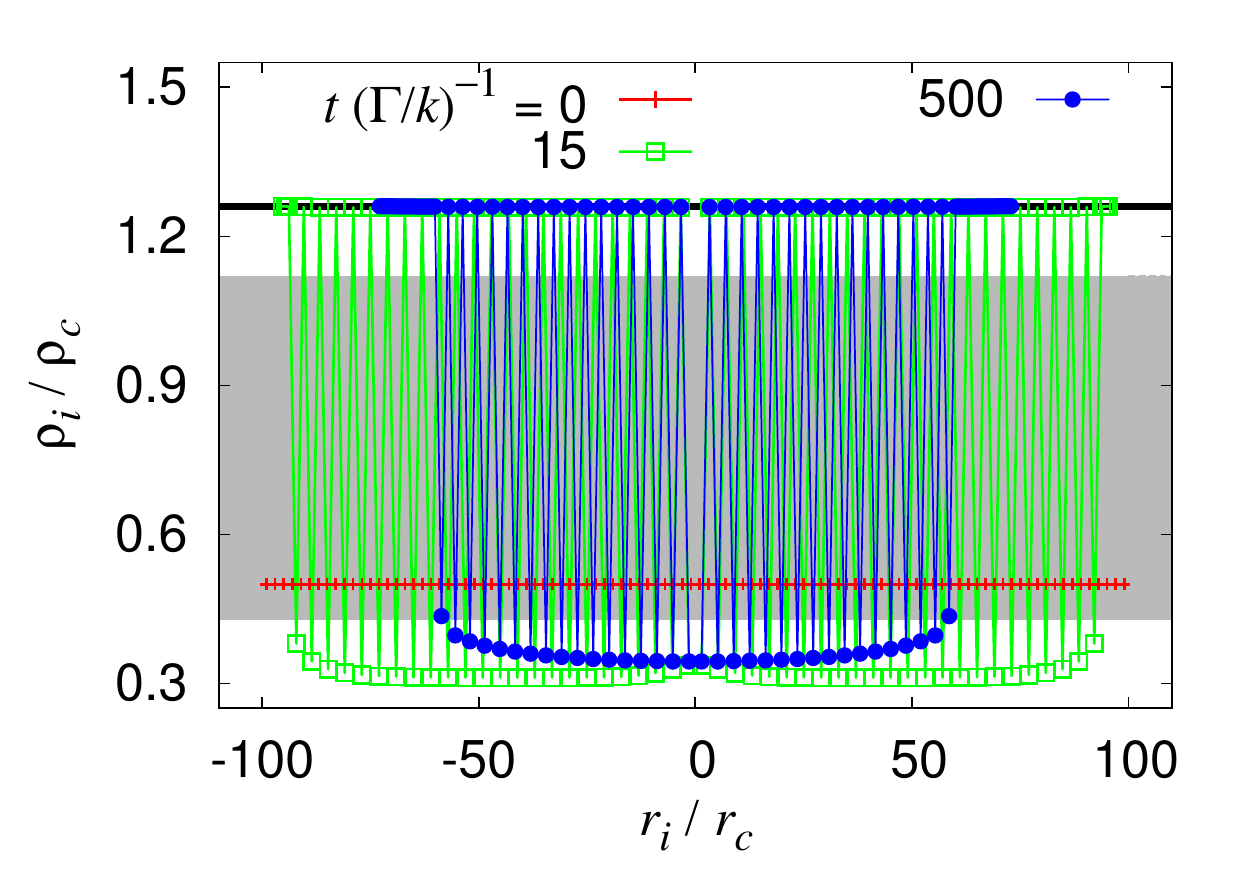}
\caption{\label{fig:scen4}
Density profiles for Scenario IV. As before, symbols, black line, and gray region indicate
the particle-resolved simulation results, the equilibrium density predicted by the theory,
and the spinodal-like interval, respectively. Here, $\rho_i^{\rm init} =0.5 \rho_c$ and $m=1.7 m_0$.
The pair formation as well as the shock-wave propagation are observed clearly.
For more details, see MOVIE IV of ESI~\cite{supple}.}
\end{figure}

%
Lastly, we describe Scenario IV. In this scenario, the initial configurations reside 
in the spinodal-like interval as $e^{(2)}(v^{\rm init}) < 0$.
Therefore, as in Scenario II, complicated configurations consisting
of touching particle pairs develop from the beginning of the dynamics. In contrast to Scenario II,
however, the density extracted from the stretched springs does neither correspond to the stable 
solution nor does it belong to the basin of the stable fixed point. Moreover, the stretched 
configurations are no longer in the spinodal-like interval, once the spinodal-like dynamics 
are settled. Therefore, one may expect a shock-wave dynamics as in Scenario III. 
Indeed, we observe once again a shock-wave propagation, see Fig.~\ref{fig:scen4}. 
In this scenario, it is the touching of the touching pairs
instead of the single particles which constitutes the dynamics of the shock wave. 
We also confirm that the numerical integration of the theory exhibits similar time evolution 
in the density distributions as shown in Appendix~\ref{ap:numerics}. However, a quantitative 
description of the shock structure/position in terms of the theory is still in progress. 

\subsection{Dynamical state diagram}
\label{sec:diagram}
\begin{figure}
\centering
\includegraphics[width=0.48\textwidth]{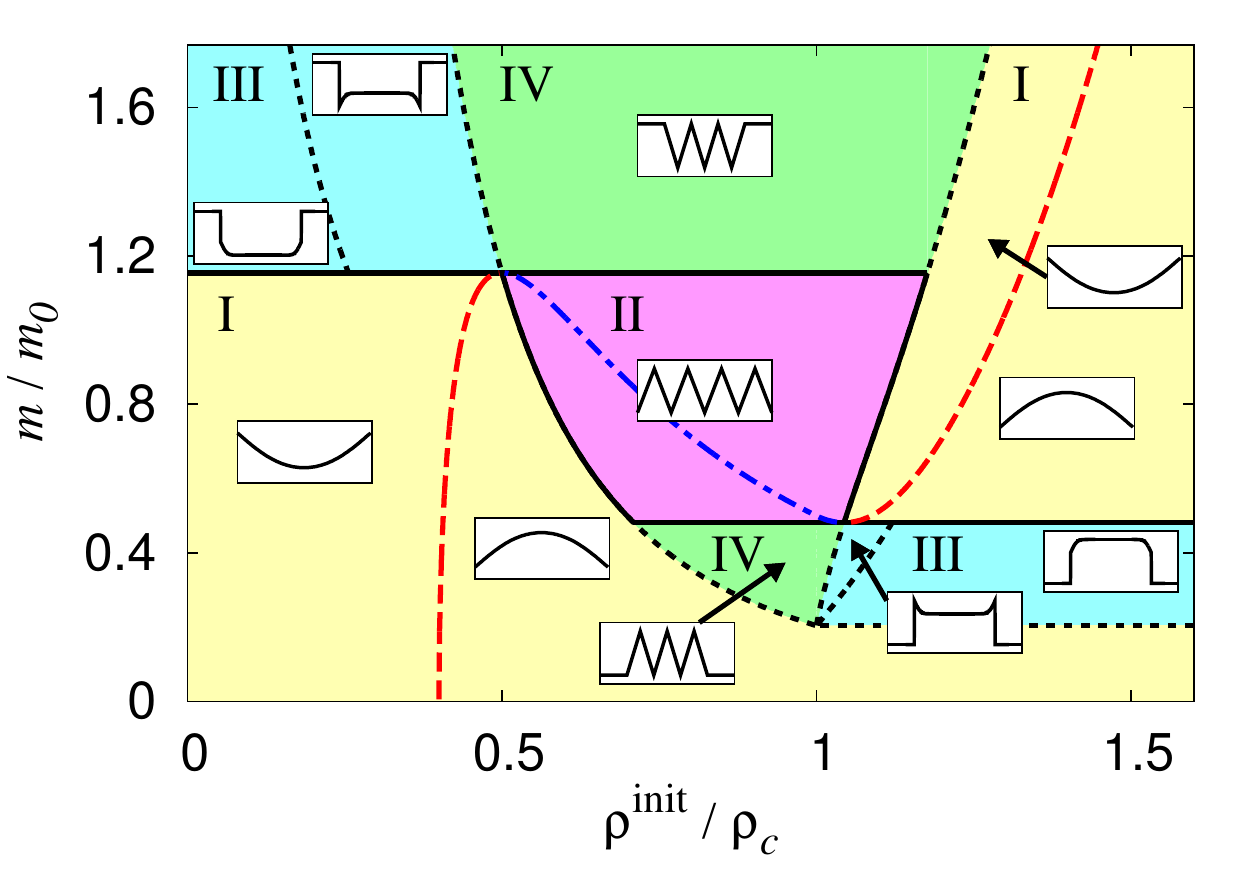}
\caption{\label{fig:diagram}
Dynamical state diagram. Black dotted lines indicate boundaries between dynamical scenarios,
while black solid lines discriminate between the states with different equilibrium configurations,
see Fig.~\ref{fig:landscape}. Red dashed lines represent the equilibrium states of global 
energetic minima while the blue dotted-dashed line corresponds to the unstable local maximum 
points in the pairwise energy. We note that the black solid lines between yellow 
and magenta areas and the black dotted lines between green and cyan regimes 
and between green and yellow areas constitute the spinodal line with $e^{(2)}(\rho) =0$. 
Additional black dotted lines inside cyan regions
are of $\rho_+/\rho_c$ for $m/m_0 \gtrsim 1.15$ and of $\rho_-/\rho_c$ for 
$0.21 \lesssim m/m_0 \lesssim 0.48$. 
We also note that green and magenta regions are the spinodal-like intervals while yellow regions 
correspond to the basins of attraction of the equilibrium points. The same colors
as in Fig.~\ref{fig:landscape} are used to identify the different dynamical scenarios. 
Schematic density profiles of corresponding dynamical scenarios are indicated.}
\end{figure}

%
Putting together the four different scenarios, we present in Fig.~\ref{fig:diagram} 
a dynamical state diagram of the one-dimensional dipole-spring model. 
No other qualitatively different scenario was found for the present
energy with at most two equilibrium points. Schematic figures 
represent the density profiles at intermediate time scales after the settlement of 
the initial pair-formation dynamics but before full equilibration. 
Here, let us elucidate the observed phenomena. 

When the magnetic moment is very small 
($m/m_0 \lesssim 0.21$), the effects of the magnetic interactions 
are negligible and the touching/detachment dynamics does not play a significant role.
If we consider the regime of strong magnetic moments ($m/m_0 \gtrsim 1.15$), 
the magnetic interactions significantly affect the overall dynamics. 
As the magnetic interactions are strong, it is the touching of particles separated 
in the low initial density regimes (cyan) that triggers the shock-wave dynamics. 
Here, we further note that, phenomenologically, the contraction of the chain is mainly
governed by this shock-wave dynamics. 

In the case of the spinodal-like mechanism (green and magenta), the pair formation rather 
contributes to the redistribution of particles and sometimes even causes an increase of the 
chain length. Here, the dissipation of energy is faster during the initial stage of the 
pair formation than during the shock-wave propagation. This seems plausible 
as the instabilities are localized only in the vicinity of the interfaces in the case of 
the shock-wave propagation dynamics, while they are distributed across the whole system
in the spinodal-like case, simultaneously contributing to the energy dissipation during 
the pair formation. 

Opposite phenomena are observed in the range of $0.21 \lesssim m/m_0 \lesssim 0.40$. 
Even though the magnetic interactions play a significant role in this regime, 
what we observe is mostly the separation of particles as the magnetic interactions are still 
weak in this case. Specifically, we observe separation of magnetic particles starting from 
the boundaries and propagations of sharp interfaces extending expanded regions of the chain
from the left/right ends to the center in the high-density regime (cyan). 
Similarly, in the intermediate-density regime (green), the separation of particles that 
form pairs due to the spinodal-like mechanism in the initial stage of the dynamics
underlies the shock-wave propagation.

In the intermediate $m$-regime, at $0.48 \lesssim m/m_0 \lesssim 1.15$ (magenta), 
we observe heterogeneous configurations as resulting equilibrium states due to the bistability of 
the energy. However, if the dynamical theory based on the quasi-continuum equation of motion 
Eq.~\eqref{eq:regularization} is evaluated, we are not able to describe the 
emergence of this Scenario II. Only the relaxation dynamics to the global minimum 
states are found. 

\subsection{General discussion}
Lastly, we qualitatively discuss our results in comparison to general aspects of the dynamics 
of phase separation. First, in the present case, it is found that the boundaries initiate 
the dynamics of the systems, instead of thermal fluctuations as for general scenarios of 
phase separation. Secondly, the growth mechanism of touching clusters (or their separation dynamics)
following the spinodal-like initial dynamics is different from the phase
separation due to different conservation laws. 
In sharp contrast to scenarios of typical phase separation, in our case the overall size of the 
system may change over time. Thus the particle number is conserved in the dipole-spring system, 
instead of the global density as in typical scenarios of phase separation. This 
counteracts the coexistence of two phases of different densities but rather promotes 
the transition to only one phase. Consequently, the shock-wave propagation dominates the long-time 
relaxation dynamics of the system, driving the change in extension of the chain and promoting
the overall transformation of the whole system.

Apart from that, as a technical detail, the underlying background of the regularization 
is also different in our quasi-continuum description. While, for instance, the interface 
itself contributes to the free energy in the Cahn-Hilliard equation in the form of 
gradient terms~\cite{Cahn1958, Pego1989}, it is only the discreteness of the system 
that gives rise to the regularization in our case. We stress, however, that 
the spinodal-type mechanism based on the structure of the underlying energy is 
formally rather analogous, leading to the emergence of pair/cluster formation. 

Altogether, the touching/detachment dynamics can be related to a spinodal-type
mechanism, while the interfacial shock-wave propagation governing the long-time dynamics
in certain cases may rather be comparable to a scenario of domain growth.
Different scenarios of touching/detachment dynamics are summarized in Table~\ref{table}.

\begin{table*}
\begin{tabular}{c|c|c|c}
\hline
\hline
Scenario & $\rho_i (r_i)$ & Intermediate configuration & Equilibrium state \\
\hline
II & \parbox[c]{5em}{\includegraphics[width=1.5cm]{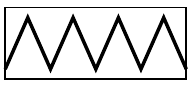}} 
	& \parbox[c]{18em}{\includegraphics[width=5cm]{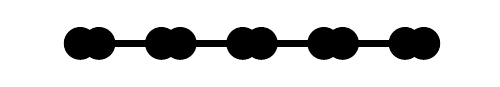}} 
	& Heterogeneous \\
\hline
III & \parbox[c]{5em}{\includegraphics[width=1.5cm]{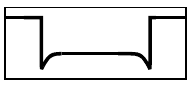}} 
		\parbox[c]{5em}{\includegraphics[width=1.5cm]{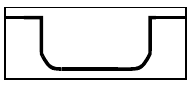}} 
	& \parbox[c]{18em}{\includegraphics[width=5cm]{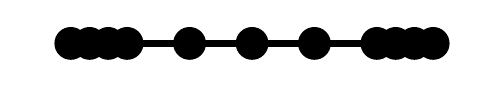}} 
	& Uniform, $\rho > \rho_c$ \\
 	& \parbox[c]{5em}{\includegraphics[width=1.5cm]{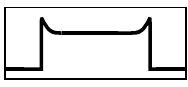}} 
		\parbox[c]{5em}{\includegraphics[width=1.5cm]{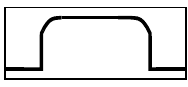}} 
	& \parbox[c]{18em}{\includegraphics[width=5cm]{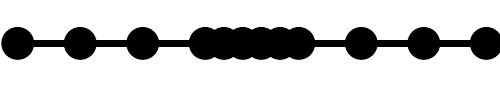}}  
	& Uniform, $\rho < \rho_c$ \\
\hline
IV & \parbox[c]{5em}{\includegraphics[width=1.5cm]{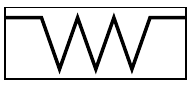}}  
	& \parbox[c]{18em}{\includegraphics[width=5cm]{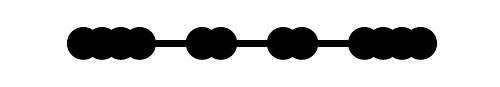}} 
	& Uniform, $\rho > \rho_c$ \\
 	& \parbox[c]{5em}{\includegraphics[width=1.5cm]{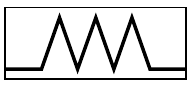}} 
	& \parbox[c]{18em}{\includegraphics[width=5cm]{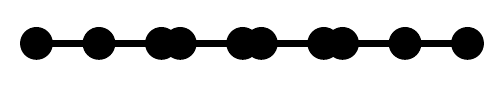}} 
	& Uniform, $\rho < \rho_c$ \\
\hline
\hline
\end{tabular}
\caption{\label{table}
Schematic graphical representations of the scenarios involving the touching/detachment dynamics.
While schematic figures for the intermediate density profiles are displayed in the second 
column, corresponding intermediate configurations during the relaxation to stationary equilibrium 
states are briefly portrayed in the third column. Values of $m$ and $\rho^{\rm init}$ can be
identified from the dynamical state diagram in Fig.~\ref{fig:diagram} by comparing the 
corresponding schematic plots in the second column. In the last column, the characteristics 
of their final stationary equilibrium configurations are summarized. In all of the three 
scenarios, a spinodal-type mechanism underlies the initial touching and detachment dynamics 
of the magnetic particles. However, the long-time dynamics are always dominated by the 
shock-wave propagation, except for Scenario II, in which there is no further long-time 
relaxation dynamics.}
\end{table*}

\section{Summary and outlook}
\label{sec:discussion}
%
Until now, we have investigated the relaxation dynamics of a one-dimensional dipole-spring model.
We have revealed that a type of spinodal decomposition mechanism plays a central role
in the touching or detachment dynamics of magnetic particles and that shock-wave-type propagations
can dominate the long-time relaxation dynamics to the equilibrium states. The boundary effects 
are shown to be an essential ingredient for the initiation and the subsequent qualitative 
appearance of the dynamics, while the discreteness of the system regularizes the 
continuum equation of motion.
It is remarkable that even these simple one-dimensional systems exhibit heterogeneous
scenarios in spite of the homogeneity in initial and, mostly, equilibrium configurations.
A variety of rich dynamics involves the interplay between the formation of particle pairs
and the shock-wave propagation. 

There still remains plenty of space for further extensions of the present study.
First of all, the response of the system to time-dependent magnetic fields
is of interest. Specifically, effects of the touching/detaching dynamics on the dynamic moduli 
of the system~\cite{Pessot2016,Pessot2018} may deepen the understanding of the magneto-mechanical 
couplings in ferrogels. 
Extensions of the model to two- and three-dimensional systems are also an important step.
In part, we anticipate similar dynamics for strong directed magnetic interactions,
as then, likewise, one-dimensional chain-like aggregations will form aligned along the direction 
of an applied external magnetic field~\cite{Pessot2016,Gundermann2017,Schumann2017,Pessot2018}. 
Already, our one-dimensional simulation results suggest that the global minimum states 
in the intermediate regime could be non-uniform. Even more possibilities arise in two or 
three dimensions and, therefore, even richer dynamics are expected to be observed. 
In addition to that, effects of thermal fluctuations should be clarified as well~\cite{Cremer2017}.
For example, if heterogeneous initial configurations are taken into account, 
we observe the onset of the shock-wave propagation in the particle-resolved simulation
for long-range magnetic interactions even from the interior of the chain. 
One may expect similar phenomena in the system induced by thermal fluctuations, 
which may correspond to the nucleation of dense clusters or soft components.

We expect that the results discussed in this study can be confirmed from experiments.
Indeed, the experimental technology these days enables researchers to capture the configuration 
at a certain time point~\cite{Gundermann2017} or to provide a temporal resolution of the dynamics
of corresponding systems~\cite{Puljiz2016, Huang2016}. Therefore, supported by quantitative 
analysis of the data, the formation of particle pairs and the propagation of sharp interfaces 
might be verified. Still, there is a possibility that the imperfections inherent in experimental 
samples may obscure such verification. However, there are efforts to construct uniform 
nanocomposite samples~\cite{Feld2017}. With the aid of such an approach, the rigorous 
comparison between theory and experiments could be achieved.

Meanwhile, especially in interpreting possible experimental results, randomness in the network
connectivity as well as in the arrangement and size of the magnetic particles should be 
taken into account. 
Still, one may expect a similar dynamics, consisting of pair formation and
shock-wave propagation. For example, touching pairs and compact chain formation are observed 
even in three-dimensional inhomogeneous dipole-spring systems based on experimentally observed 
particle configurations~\cite{Pessot2018}. 
However, details such as the size of the touching clusters or the initiation mechanism of 
exit
the dynamics may differ. If heterogeneity is introduced in the spring constant, 
softer parts of the chain may form a touching cluster more easily than other parts of the system 
and, therefore, the chain formation dynamics could be initiated in various parts of the system. 
In this case, we speculate that a kind of coupling between the interfaces may
play a certain role. Verification of such couplings could be a challenging task in  
theoretical as well as in experimental studies.

In short, we expect that our results may serve as an essential building block in understanding
the dynamics of more realistic models for ferrogels. However, we also note that a 
further adjusted continuum theory with fine-tuned regularization terms should be 
devised to fully describe the whole dynamics, especially in the bistable regime.
This is left for future works.

\section*{Acknowledgments}
We thank Giorgio Pessot for providing codes which were useful for the initiation of this study. 
We also thank Giorgio Pessot, Peet Cremer, J\"urgen Horbach, and Benno Liebchen for helpful 
discussions and comments. This work was supported by funding 
from the Alexander von Humboldt Foundation (S.G.) and
from the Deutsche Forschungsgemeinschaft through the priority program SPP 1681,
grant nos. ME 3571/3 (A.M.M) and LO 418/16 (H.L.).
%
\onecolumngrid
\appendix

\section{Equations of motion for the particles}
\label{ap:eq_motion}
%
We describe the equations of motion for the particles in the magnetic chain in detail. 
The equations for the boundary particles are shown explicitly as well. 

Obviously, the term on the right-hand side of Eq.~\eqref{eq:overdamped} consists of three parts.
The first one of them, resulting from the elastic energy, reads 
\begin{align}
-\frac{\partial E_{el}}{\partial r_i} &= k(r_{i+1}-r_i) -k(r_i -r_{i-1})
\end{align}
for $i=2,\ldots,N$ and
\begin{align}
-\frac{\partial E_{el}}{\partial r_1}= k(r_2 -r_1-a), \quad
-\frac{\partial E_{el}}{\partial r_{N+1}} = -k(r_{N+1} -r_{N}-a).
\end{align}

Second, the contributions from the magnetic dipole-dipole interaction take the form
\begin{align}
\label{eq:f_m_in}
-\frac{\partial E_{m}}{\partial r_i} &=\frac{3 \mu_0 b m^2}{4\pi} 
	\left[ -\sum_{j>i} \frac{1}{\left( r_j -r_i \right)^4}
		+\sum_{j<i} \frac{1}{\left( r_j -r_i \right)^4} \right]
\end{align}
for $i=2,\ldots,N$, and 
\begin{align}
\label{eq:f_m_bound}
-\frac{\partial E_{m}}{\partial r_1} =-\frac{3\mu_0 b m^2}{4\pi} 
	\sum_{j=2}^{N+1} \frac{1}{\left( r_j -r_1 \right)^4}, \quad 
-\frac{\partial E_{m}}{\partial r_{N+1}} =\frac{3\mu_0 b m^2}{4\pi} 
	\sum_{j=1}^{N} \frac{1}{\left( r_j -r_{N+1} \right)^4}.
\end{align}
We note that the nearest-neighbor magnetic dipole-dipole interaction is obtained
from the above equations by constraining the summations in $j$ to nearest neighbors. 

Lastly, we have
\begin{align}
-\frac{\partial E_{st}}{\partial r_i} &= {f}_{st} (r_{i,i+1}) -{f}_{st} (r_{i-1,i})
\end{align}
for $i=2,\ldots,N$ and
\begin{align}
-\frac{\partial E_{st}}{\partial r_1} &= {f}_{st} (r_{1,2}), \quad
-\frac{\partial E_{st}}{\partial r_{N+1}} = -{f}_{st} (r_{N,N+1}) 
\end{align}
from the steric repulsion energy, where
\begin{align}
f_{st} (r) &\equiv \epsilon^s \Theta (r-r_c)
	\left[ -\frac{12}{\sigma^s} \left( \frac{r}{\sigma^s} \right)^{-13}
	+\frac{6}{\sigma^s} \left( \frac{r}{\sigma^s} \right)^{-7} -{c^s (r-r_c)} \right].
\end{align}

\section{Singular perturbation theory}
\label{ap:singular}
%
We define $v \equiv r_x$ and rescale the time variable by introducing $\tau \equiv (\Delta x)^2\,t$
for convenience. Then Eq.~\eqref{eq:regularization} reads
\begin{align} \label{eq:sing_pert}
\Gamma v_\tau = \left[  e^{(1)} (v) 	+(\Delta x)^2 \left( \frac{1}{12} e^{(2)} (v)\, v_{xx} 
	+\frac{1}{24} e^{(3)} (v)\, (v_{x})^2 \right) \right]_{xx}.
\end{align}

\begin{figure*}
\centering
\includegraphics[width=0.48\textwidth]{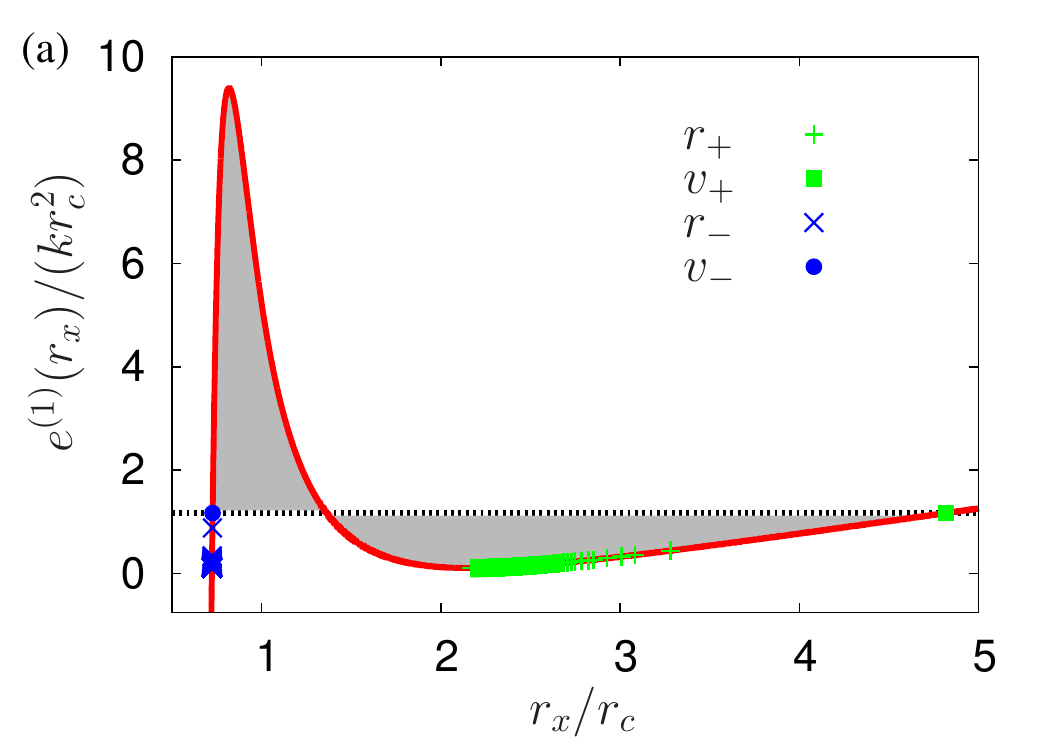}
\includegraphics[width=0.48\textwidth]{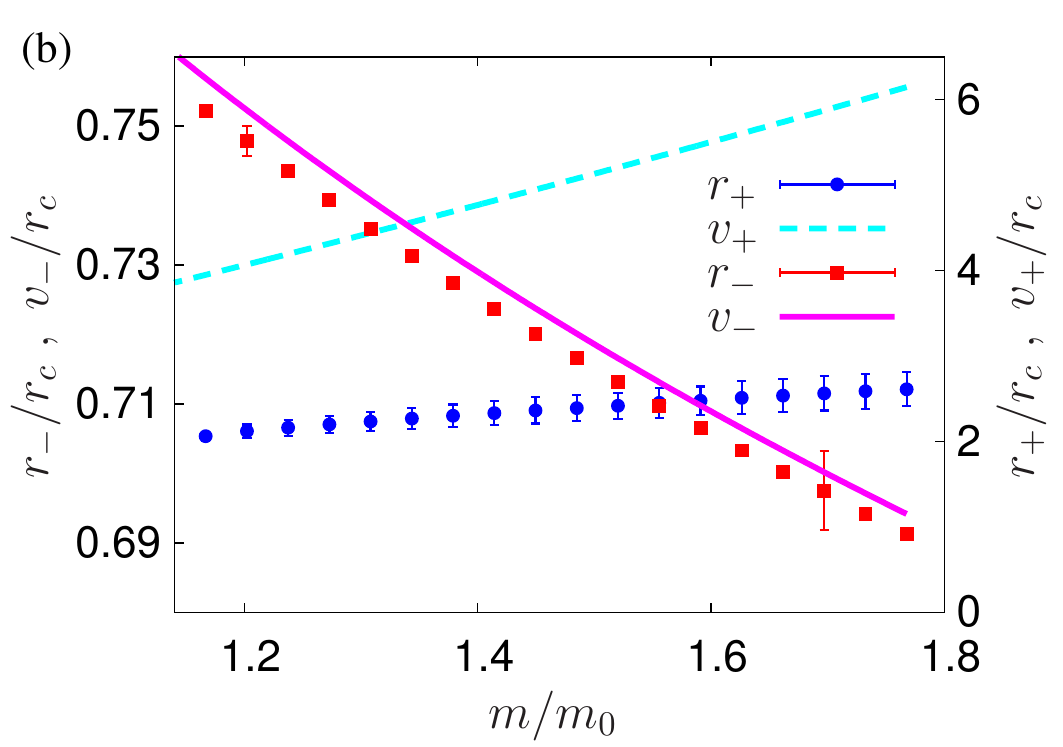}
\includegraphics[width=0.48\textwidth]{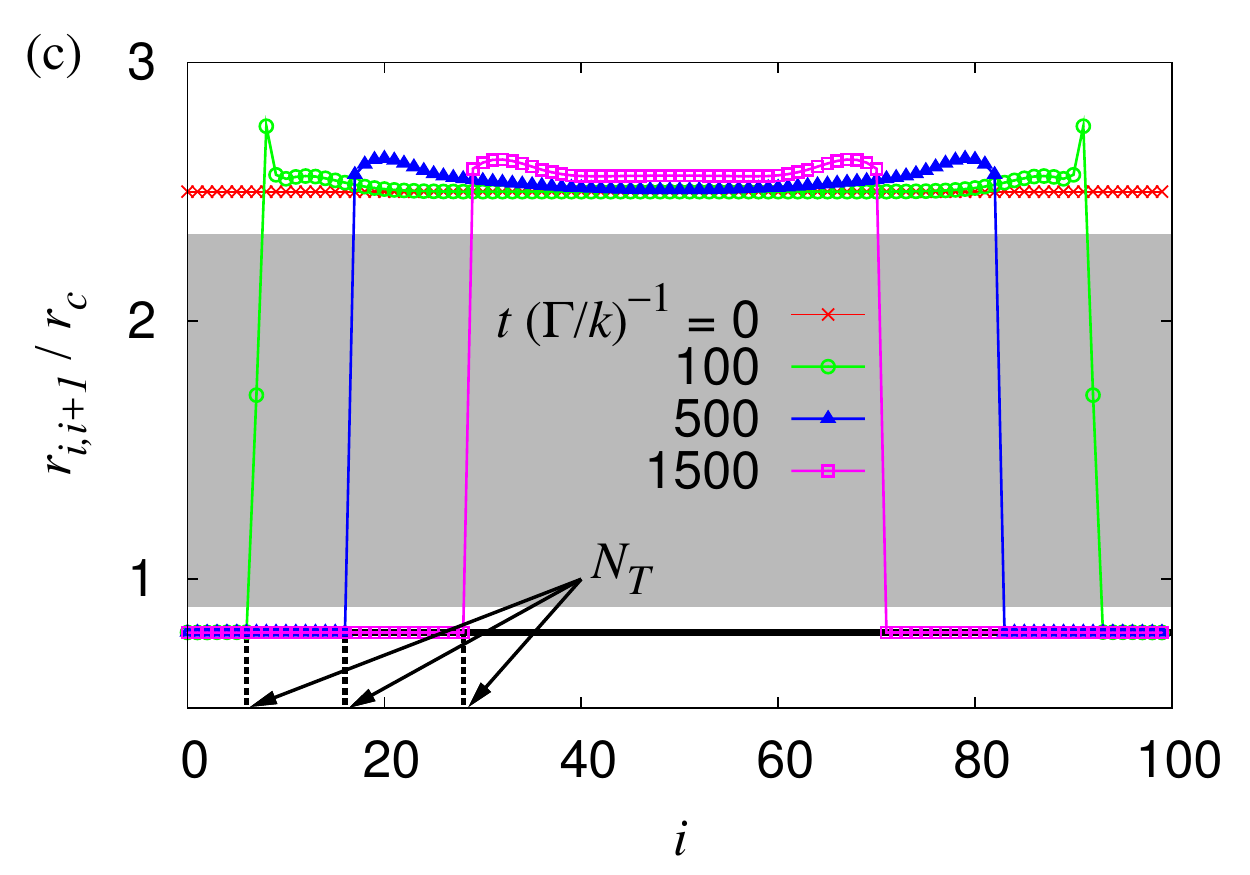}
\includegraphics[width=0.48\textwidth]{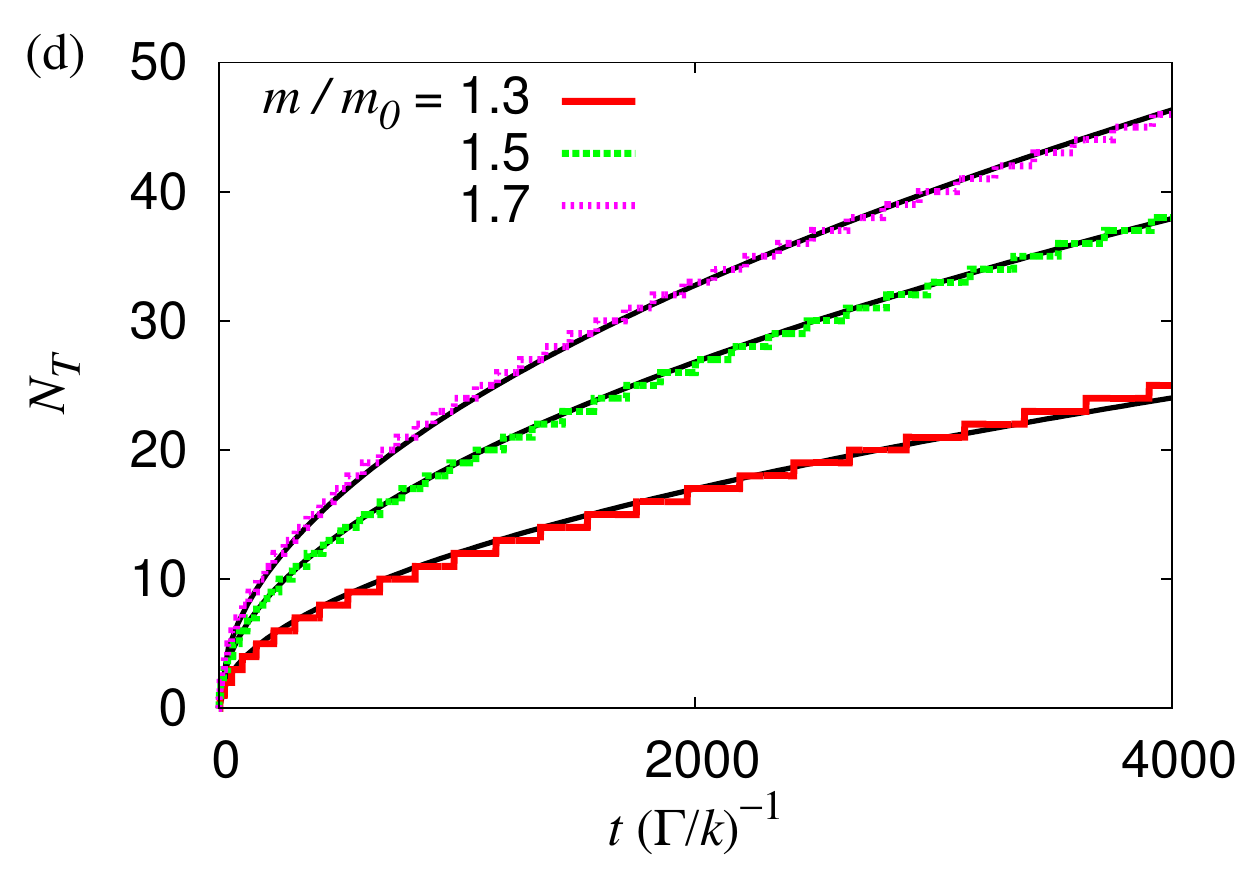}
\caption{\label{fig:shock}
(a) How the equal area rule applies in practice is described: the structures of the shocks are
given by the filled blue circle ($v_-$) and the filled green square ($v_+$),
corresponding to the intersection points between the red line representing $e^{(1)}$ and 
the black dotted line, 
where $v_-$ and $v_+$ are the values of $r_x$ behind and in front of the shock. 
The gray areas above and below the black dotted line need to be equal to each other. 
Plus and cross symbols representing the particle-resolved simulation results are displayed 
as well. Specifically, $r_-$ and $r_+$ are given as the lengths of the springs 
in front of and behind the abrupt jump in the density, respectively.
(b) The values of $v_+$ and $v_-$ (magenta and cyan solid lines) computed 
from Eqs.~\eqref{singular_sol1} and~\eqref{singular_sol2} are compared with the particle-resolved 
simulation results, $r_+$ and $r_-$ (blue circles and red squares).
We note that the values on the left axis correspond to those of $r_-$ (red squares) and 
$v_-$ (magenta solid line), while $r_+$ (blue circles) and $v_+$ (cyan dashed line) are 
represented by the axis on the right.
(c) Fig.~\ref{fig:scen3}(a) for Scenario III is replotted in terms of $r_{i,i+1}$ 
as a function of $i$. The values of $N_T$ are explicitly depicted by arrows. 
(d) The values of $N_T$ as functions of time $t$ are plotted for $m=1.3 m_0$, $1.5 m_0$, and $1.7 m_0$
with $\rho_i^{\rm init} =0.4 \rho_c$. We extract the values of $x_s$ by numerically fitting the 
results to the relation $N_T =x_s \sqrt{t}$ (black solid lines), which here are found to be 
$x_s (\Gamma/k)^{1/2} =$ 0.380 ($m/m_0 = 1.3$), 0.600 ($m/m_0=1.5$), and 0.732 ($m/m_0 = 1.7$).}
\end{figure*}

Introducing the extended variable $\displaystyle \tilde{x}=\frac{x-s(\tau)}{(\Delta x)}$,
we probe an interlayer solution which describes the behavior of the system in the vicinity 
of the shock front at $s(\tau)$. Under a change of variables $v = \tilde{v} (\tilde{x},\tau)$ 
with $\displaystyle v_x = \frac{1}{\Delta x} \tilde{v}_{\tilde{x}}$
and $\displaystyle v_{\tau}= \tilde{v}_{\tau}
-\frac{\dot{s}(\tau)}{\Delta x} \tilde{v}_{\tilde{x}}$,
Eq.~\eqref{eq:sing_pert}, to the leading order of $\Delta x$, becomes
\begin{align}
\left(  e^{(1)} (\tilde{v}) +\frac{1}{12}e^{(2)} (\tilde{v})\, \tilde{v}_{\tilde{x} \tilde{x}} 
+\frac{1}{24} e^{(3)}(\tilde{v})\, (\tilde{v}_{\tilde{x}})^2 \right)_{\tilde{x}\tilde{x}} = 0,
\end{align}
which has a solution of the form
\begin{align}
e^{(1)} (\tilde{v}) +\frac{1}{12}e^{(2)} (\tilde{v})\, \tilde{v}_{\tilde{x}\tilde{x}}
+\frac{1}{24} e^{(3)} (\tilde{v})\, (\tilde{v}_{\tilde{x}})^2 = A\tilde{x}+B.
\end{align}
For the interlayer solutions $\tilde{v}$ and $\tilde{v}_{\tilde{x}}$ must approach 
constant values $v_\pm$ and $0$ as $\tilde{x} \to \pm \infty$ and, therefore, 
\begin{align}
\label{singular_sol1}
A=0,\quad B = e^{(1)}(v_-)=e^{(1)}(v_+).
\end{align}
Further multiplying by $\tilde{v}_{\tilde{x}}$, we also find
\begin{align}
\left( B\tilde{v} \right)_{\tilde{x}}
= \left( e (\tilde{v}) +\frac{1}{24} e^{(2)}(\tilde{v})\, (\tilde{v}_{\tilde{x}})^2 \right)_{\tilde{x}},
\end{align}
which leads to the equal-area rule~\cite{Pego1989,Witelski1996} for the interlayer solution 
of the form 
\begin{align}
\label{singular_sol2}
B(v_+ -v_-)=e(v_+)-e(v_-).
\end{align}
Numerically solving Eqs.~\eqref{singular_sol1} and~\eqref{singular_sol2}, 
one can compute the values of $v_+$ and $v_-$, which specify the structure of the shocks.
The construction of the equal area rule and the resultant shock structures are presented 
in Figs.~\ref{fig:shock}(a) and (b).

We turn to the propagation speed of the shock-wave front $s(\tau)$. As our equation of motion in 
the leading order takes the form of a diffusion equation, we consider a similarity solution 
$U(z) =v(x,\tau)$ with the similarity variable $z\equiv x/\sqrt{\tau}$.
The equation of motion in the leading order follows as
\begin{align} \label{eq:diff_v_tau}
\Gamma v_\tau = (e^{(2)}(v)\,v_x )_x 
\end{align}
in terms of $x$ and $\tau$. It can be rewritten as
\begin{align} \label{eq:similarity}
\frac{\Gamma}{2}z \frac{\dd U(z)}{\dd z}
	=-\frac{\dd}{\dd z} \left( e^{(2)}(U) \frac{\dd U(z)}{\dd z} \right)
\end{align}
in terms of the similarity variable $z$. In particular, the quantity of interest is the 
coefficient $x_s$ which is defined by a similarity relation 
$s(\tau) =x_s \sqrt{\tau}$. Quantifying the values of $x_s$ with the Whitham's 
derivation~\cite{Whitham1999}, one can describe the dynamics of the shock-wave propagation.
For self-containedness, we briefly summarize the procedure, following 
Refs.~\onlinecite{Witelski1995} and~\onlinecite{Whitham1999}. 
We also note that an equivalent result was obtained~\cite{Pego1989} 
with the aid of mathematical consideration of Stephan problems~\cite{Crank1975}.

First, we consider the diffusion flux $q$ in a region $x_1 > x >x_2$ where a balance 
between the net inflow across $x_1$ and $x_2$ in a region is described by
\begin{align} \label{eq:net_flux}
\Gamma \frac{\dd}{\dd \tau} \int_{x_1}^{x_2} v(x,\tau)\,\dd x
+ q (x_1,\tau)-q(x_2, \tau)=0,
\end{align}
leading to the conservation form
\begin{equation}
\Gamma \frac{\partial v}{\partial \tau}+\frac{\partial q}{\partial x} =0.
\end{equation}
Therefore, we define the diffusion flux 
as $q\equiv -e^{(2)}(v)\,v_x$ [see Eq.~\eqref{eq:diff_v_tau}].
We then extend the above consideration to a case with a discontinuity at $x=s(\tau)$.
In this case, Eq.~\eqref{eq:net_flux} can be rewritten as follows~\cite{Whitham1999}:
\begin{align}
q(x_1,\tau)-q(x_2,\tau)
	=&\Gamma \frac{\dd}{\dd \tau}\int_{x_2}^{s(\tau)}v(x,\tau)\dd x
	+\Gamma \frac{\dd}{\dd \tau}\int_{s(\tau)}^{x_1}v(x,\tau)\dd x \\
=&\Gamma v(s^-,\tau)\frac{\dd s}{\dd \tau}
	+\int_{x_2}^{s(\tau)} \Gamma v_{\tau} (x,\tau) \dd x 
    -\Gamma v(s^+,\tau)\frac{\dd s}{\dd \tau}
	+\int_{s(\tau)}^{x_1} \Gamma v_{\tau} (x,\tau)\dd x.
\end{align}
With the limits $x_1 \to s^+$ from above and $x_2 \to s^-$ from below, we obtain
\begin{align}
\Gamma \frac{\dd s}{\dd \tau} = \frac{q(s^-,\tau)-q(s^+,\tau)}
	{v(s^-,\tau)-v(s^+,\tau)} \equiv \frac{[q]}{[v]},
\end{align}
where square brackets denote the jump of the contained value across the interface.
For the similarity solution, $q(x,\tau)=-\tau^{-1/2}e^{(2)}(U)U_z$, and therefore
the above equation is cast into the form~\cite{Witelski1995}
\begin{align}
\label{eq:shock_position}
\frac{1}{2}x_s = -\frac{[e^{(2)}(U) U_z]}{[\Gamma U]},
\end{align}
which finally determines the propagation speed of the shock front. Numerically solving 
Eqs.~\eqref{eq:similarity} and~\eqref{eq:shock_position}, we obtain the values of $x_s$ which 
are presented in Fig.~\ref{fig:scen3}(b). Using these values, we can compute, for instance,
the number of touching particles $N_T(t)$ in one end. 
Specifically, scaling back to the time $t$, we have
\begin{align}
N_T (t) =\frac{s(\tau)}{\Delta x} =\frac{x_s \sqrt{(\Delta x)^2 t}}{\Delta x} 
	=x_s \sqrt{t}.
\end{align}
As expected, the number of touching particles is independent of the value of $\Delta x$. 
Predicted values of $x_s$ are shown in Fig.~\ref{fig:scen3}(b), together with those extracted from the simulations results by the procedure described in Figs.~\ref{fig:shock}(c) and (d).

\section{Numerical integration of the continuum equation of motion}
\label{ap:numerics}
%
In this appendix, we describe the algorithm used in integrating the quasi-continuum equation 
of motion. The algorithm is a modified version of the upwind scheme~\cite{Patankar1980,Hirsch2007},
which is widely used to find propagating solutions to wave equations. 
However, if it is directly applied to the magnetic chain under contraction,
for instance, the shrinkage of the chain is rather exaggerated 
as the particles behind an interface receive biased information towards
the particles in front of the interface~\cite{Patankar1980}, which may 
impose a resistance to contraction. To compensate such an artifact, we introduce an additional 
downwind-biased step and write the discretized equation for each time step $\Delta t$ as follows:
\begin{align} \label{eq:scheme}
\Gamma r &(x,t+\Delta t) = \Gamma r(x,t)+\frac{\Delta t}{2} \left[ 
	e^{(2)}\left( \frac{r(x,t)-r(x-\Delta x,t)}{\Delta x} \right) 
	\frac{r(x+\Delta x,t)+r(x-\Delta x,t)-2r(x,t)}{(\Delta x)^2} \right] \nn \\
	&+\frac{\Delta t}{2} \left[ 
	e^{(2)}\left( \frac{r\left(x+\Delta x,t+\frac{\Delta t}{2}\right)-r\left(x,t+\frac{\Delta t}{2}\right)}{\Delta x} \right)
	\frac{r\left(x+\Delta x,t+\frac{\Delta t}{2}\right)+r\left(x-\Delta x,t+\frac{\Delta t}{2}\right)-2r\left(x,t+\frac{\Delta t}{2}\right)}{(\Delta x)^2} \right]. 
\end{align}

As already pointed out in Ref.~\onlinecite{Barenblatt1993}, a certain form of regularization is 
always involved in the numerical integrations, which are indeed discrete. In the case of 
the numerical scheme discussed here, the dominant correction to the fully 
continuum equation of motion [Eq.~\eqref{eq:cont_DS_leading}] is given as 
\begin{align}
(\Delta x)^2 \left[ \frac{1}{12} e^{(2)}(r_x)\,r_{xxxx} +\frac{1}{6}e^{(3)}(r_x)\,r_{xx} r_{xxx}
+\frac{1}{8} e^{(4)}(r_x)\, (r_{xx})^3 \right].
\end{align}
Interestingly, the terms are almost equivalent to the leading order regularization in 
Eq.~\eqref{eq:regularization}. Therefore, we conclude that the algorithm discussed above provides 
solution to the continuum equation of motion but with a slightly different type of regularization. 

\begin{figure*}
\centering
\includegraphics[width=0.48\textwidth]{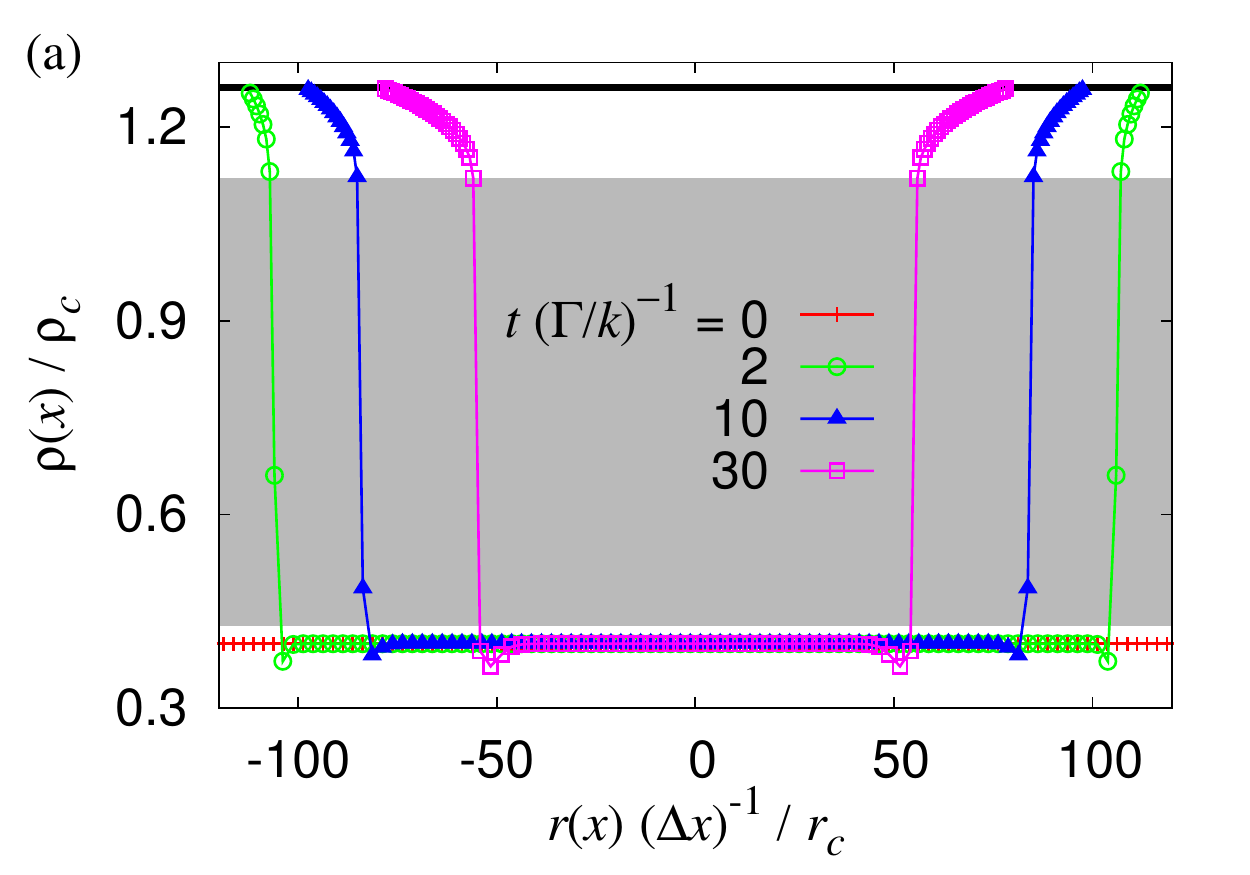}
\includegraphics[width=0.48\textwidth]{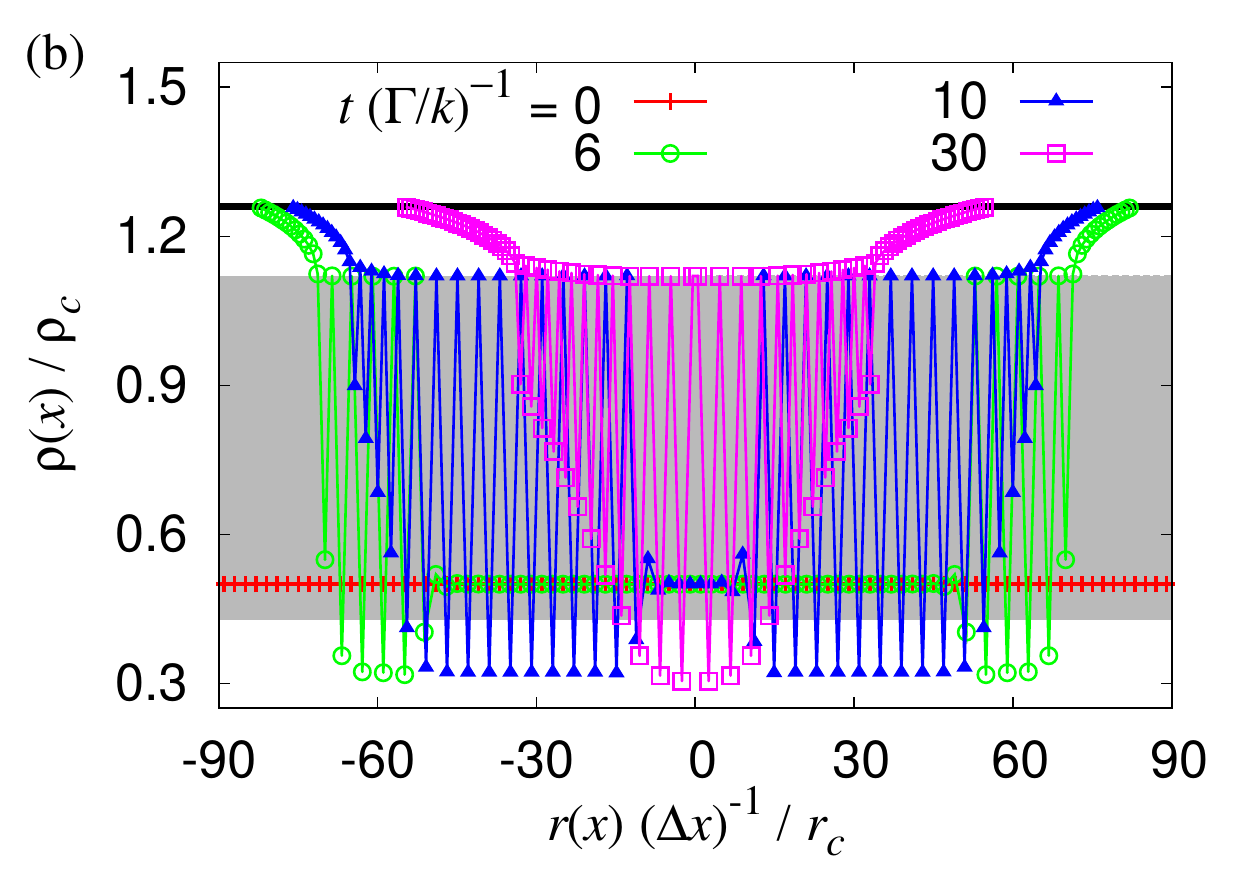}
\caption{\label{fig:numerics}
Density profiles extracted from the numerical solutions with the scheme Eq.~\eqref{eq:scheme} for
(a) Scenario III and (b) Scenario IV. In (a), $\rho^{\rm init}=0.4 \rho_c$ and $m=1.7 m_0$ have been used,
while $\rho^{\rm init}=0.5 \rho_c$ and $m=1.7 m_0$ in (b). Even though the shock waves 
obtained in this way propagate much 
faster than in the particle-resolved simulations, the essential features, including the 
spinodal-like mechanisms for the touching, pair formation, sharp-interface generation, and 
relaxation to the equilibrium state, are well verified by the numerical solutions. 
Notably, compared to Fig.~\ref{fig:scen3}(a) and Fig.~\ref{fig:scen4}, solutions with high density,
e.g., $\rho (x) \gtrsim 1.12 \rho_c$, exhibit much smoother density variations in space,
implying an excess numerical diffusion~\cite{Hirsch2007} as described in the text.}
\end{figure*}

It is well known that the upwind scheme introduces numerical diffusion of the 
interface~\cite{Hirsch2007}. The numerical integration scheme described above also seems to suffer 
from such an issue, as the numerical solutions are not consistent with Eq.~\eqref{singular_sol1}, 
which should be satisfied regardless of regularization. Specifically, it has been tested by 
plotting a figure like Fig.~\ref{fig:shock}(a) from the numerical integration results. 
Moreover, the propagation speed of the interface sensitively depends on the structure of 
the shock as manifested in Eq.~\eqref{eq:shock_position}. 
We find that the coefficient $x_s$ extracted from a 
numerical solution can be, roughly, 100 times larger 
than the one obtained by the particle-resolved simulation and 
the singular perturbation theory. Still, the essential shapes of the solutions 
agree quite well with the simulation results, as shown in Fig.~\ref{fig:numerics}.

\twocolumngrid
\bibliography{dynamics_1D_ferrogel}

\newpage

\begin{figure*}
\centering 
\includegraphics[width=1.0\textwidth]{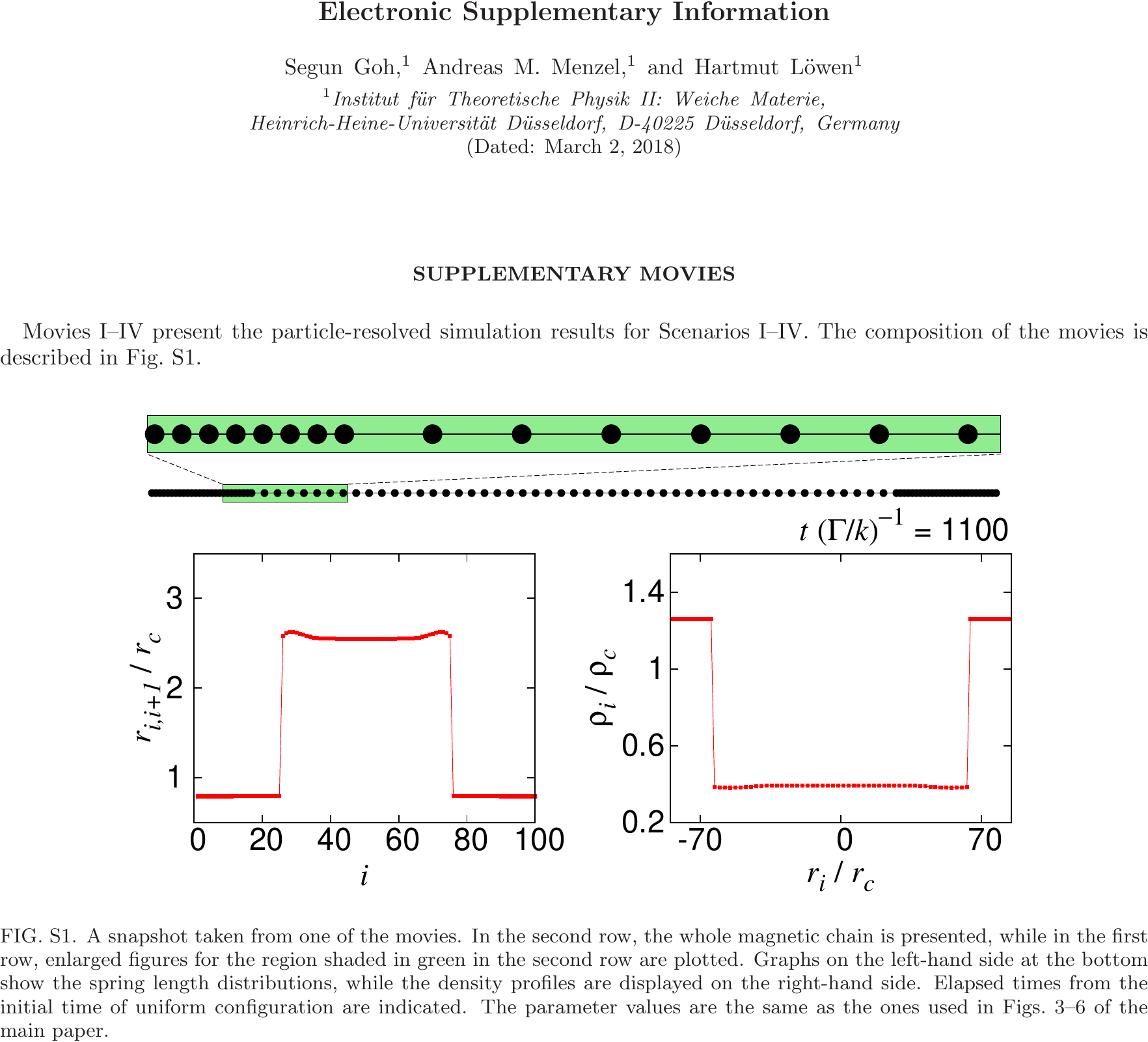}
\end{figure*}

\end{document}